\documentclass[lettersize,journal]{IEEEtran}
\usepackage{amsmath,amsfonts}
\usepackage{algorithmic}
\usepackage{algorithm}
\usepackage{array}
\usepackage[caption=false,font=normalsize,labelfont=sf,textfont=sf]{subfig}
\usepackage{textcomp}
\usepackage{stfloats}
\usepackage{url}
\usepackage{verbatim}
\usepackage{graphicx}
\usepackage{cite}
\usepackage{longtable}
\usepackage[table]{xcolor}
\usepackage{multirow}
\usepackage{tabularx}
\usepackage{booktabs}
\usepackage[normalem]{ulem}

\begin{document}

\title{Car-Following Models: A Multidisciplinary Review}

\author{Tianya Zhang, Ph.D., Peter J. Jin, Ph.D., Sean T. McQuade, Ph.D., Alexandre Bayen, Ph.D., and Benedetto Piccoli, Ph.D. 
\thanks{This work is funded by the C2SMART Project (USDOT Award \#: 69A3551747124 )}
\thanks{Manuscript received March $8^{th}$, 2024. Revised on May $18^{th}$, 2024}}



\maketitle

\begin{abstract}
Car-following (CF) algorithms are crucial components of traffic simulations and have been integrated into many production vehicles equipped with Advanced Driving Assistance Systems (ADAS). Insights from the model of car-following behavior help researchers to understand the causes of various macro phenomena that arise from interactions between pairs of vehicles. Car-following Models encompass multiple disciplines, including traffic engineering, physics, dynamic system control, cognitive science, machine learning, deep learning, and reinforcement learning. This paper presents an extensive survey highlighting the differences, complementarities, and overlaps among microscopic traffic flow and control models based on their underlying principles and design logic. It reviews a range of representative algorithms, from theory-based Kinematic Models, Psycho-Physical Models, and Adaptive Cruise Control Models to learning-based algorithms like Reinforcement Learning (RL) and Imitation Learning (IL). Additionally, it considers the impact of large Generative AI (GenAI) models, categorized as Knowledge-Driven models. The survey discusses the strengths and limitations of these models, explores their applications in various contexts, and summarizes available datasets across different domains to address knowledge gaps. By synthesizing historical developments and current trends, this survey provides a comprehensive overview of the evolution and future directions of car-following models. It highlights the importance of interdisciplinary approaches and the need for continued innovation to address emerging challenges.

\end{abstract}

\section{Introduction}
\begin{figure}[!t]
\centering
\includegraphics[width=3in]{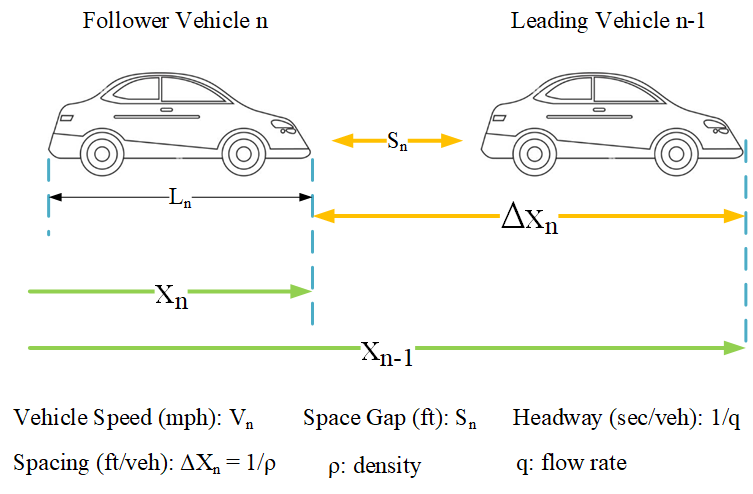}
\caption{Car-Following and Microscopic Traffic Flow Properties.}
\label{fig_1}
\end{figure}

\IEEEPARstart{T}{o} furnish a complete picture of traffic flow, vehicles on the road are studied as a whole - not merely as isolated particles. Car-following (CF) is the fundamental driving behavior that forcefully constrains the ego-vehicle from adopting a specific pattern in the presence of other vehicles (Figure \ref{fig_1}). Since the 1950s or even earlier, many researchers have focused on driver behavior modeling and applied such models to the simulation of Human Driver Vehicles (HDVs) to investigate micro- and macro-traffic phenomena \cite{1950_Reuschel,1961_Edie}. The expansion of autonomy and connectivity in the transportation system has escalated the hype for integration of car following models with Connected and Automated Vehicle (CAV) technologies, such as collaborative driving, and Vehicle-to-Everything (V2X) communication. Autonomous Vehicles (AVs) with different parameter settings or objective functions become a new source of variety and add more complexities to the existing roadway systems. The heterogeneity of vehicle types and driver behaviors gives rise to various challenges for traffic operation, which must be comprehensively analyzed before upgrading existing infrastructure. Car-following models become a powerful tool to address challenges mentioned above and answer those critical questions in the upcoming era of the connected and automated transportation systems. By analyzing the interaction between vehicles, researchers can identify efficient driving strategies that includes optimizing speeds, minimizing unnecessary acceleration and deceleration events, and maintaining optimal following distances to ensure smooth traffic flow. Modern adaptive cruise control (ACC) systems already use basic car-following principles to maintain safe distances between vehicles.    

\begin{figure*}[!htbp]
\center
\includegraphics[width=\textwidth]{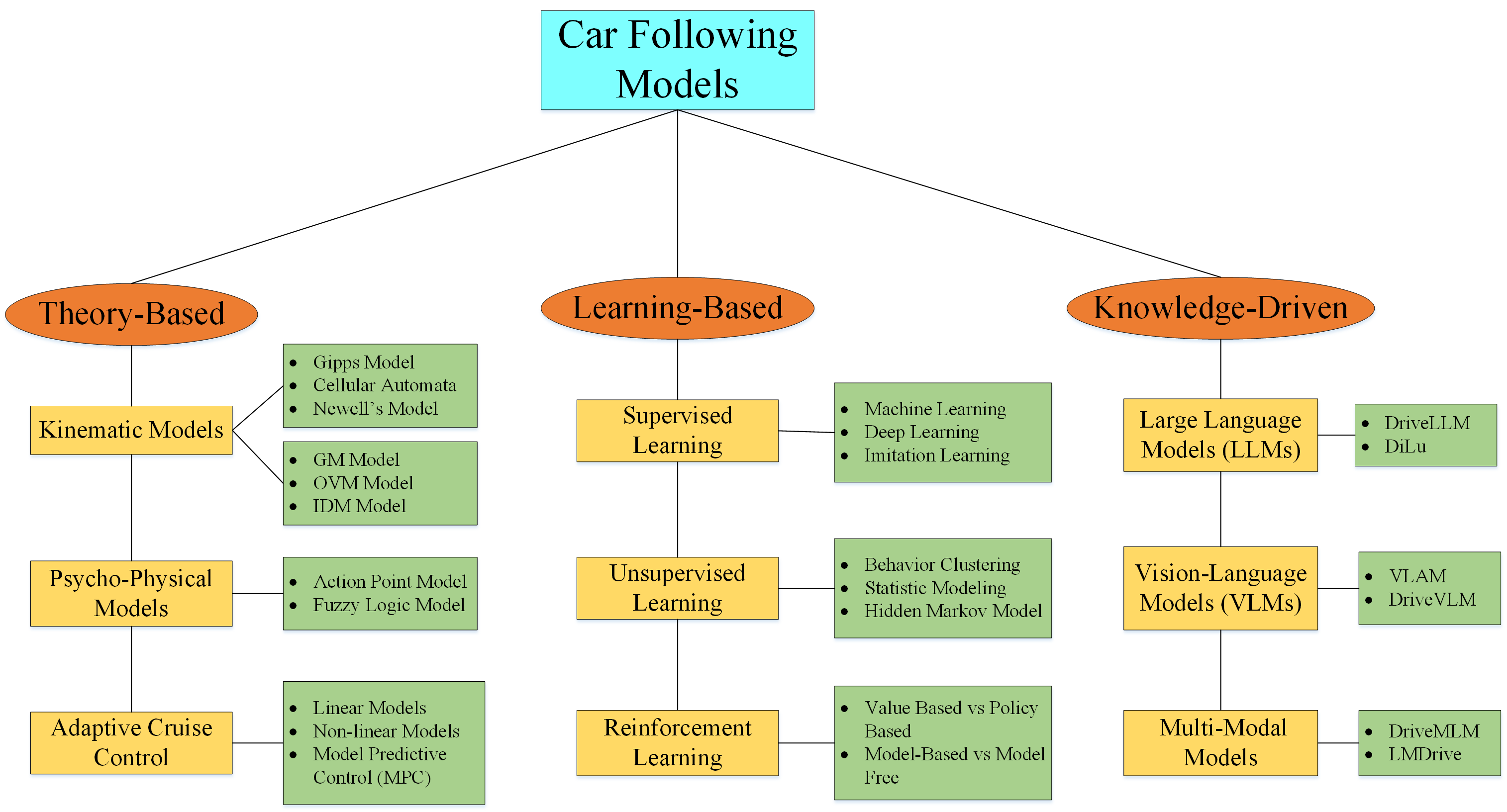}
\caption{Proposed Taxonomy of Car Following Models (CFMs) and Representative Algorithms. Kinematic Models are divided into first-order and second-order subcategories. Knowledge-driven Large GenAI Models are a revolutionary category, providing new insights and solutions with commonsense.}
\label{fig_categorization}
\end{figure*}

Existing survey papers\cite{1992_Rothery,1999_Brackstone_McDonald,2004_Olstam_Tapani,2007_Toledo,2011_Wilson_Ward, 2012_Li_Sun,2013_Treiber_Kesting,2015_Aghabayk_Sarvi_Young, 2014_Saifuzzaman_Zheng,2015_vanWageningen-Kessels_VanLint_Vuik_Hoogendoorn,2020_Matcha_Namasivayam, 2021_Ahmed_Huang_Lu, 2021_Han_Shi_Chen, 2022_Han_Wang_Wang, 2023_Sadid_Antoniou, 2023_He_Zhou_Wang, 2023_Wang_Shi_Tong_Gu_Cheng} lack of structured framework and often have limitations in terms of scope and perspective (see Table \ref{tab:car_following_review_papers}). For example, a frequently referenced paper \cite{1999_Brackstone_McDonald} mainly surveyed from traffic psychologic perspective and did not consider Intelligent Driver Model (IDM) and Optimal Velocity Model (OVM). Some surveys \cite{2015_vanWageningen-Kessels_VanLint_Vuik_Hoogendoorn, 2013_Treiber_Kesting} reviewed multiscale traffic flow models from the engineering and physics perspective, while many control theory models were omitted. Other literature reviews \cite{2012_Li_Sun, 2014_Saifuzzaman_Zheng} are limited to theory-based models and inadequate to reflect AI/ML models. By taking into account various domains and angles, this review paper aims to provide a more structured categorization of the car following models (see Figure \ref{fig_categorization}), addressing the potential gaps left by previous surveys with a narrower focus. A multidisciplinary review can be particularly beneficial in fields like intelligent vehicle technology, where insights from different disciplines – such as electrical \& computer engineering, mechanic engineering, traffic flow theory, physics, and optimal control – are crucial for developing a complete picture of how these agents operate and interact with their environment. This review paper not only integrates diverse knowledge but also promotes the discovery of innovative solutions that might be overlooked when considering a single perspective. Our contributions can be itemized as follows:

\begin{itemize}
  \item \textbf{Multidisciplinary Perspective}: Unlike previous surveys that focus on a single aspect of car-following models (e.g., kinematic characteristics, specific applications, or human factors), our paper provides a comprehensive review from multiple disciplinary perspectives. This broader approach allows for a more thorough evaluation of the state-of-the-art, challenges, and future directions of car-following models, making a unique contribution to the field.
  
  \item \textbf{Novel Taxonomy for Car-Following Models}: We introduce an effective taxonomy to categorize the broad spectrum of car-following model research. Our taxonomy is not just a listing; it delves into the underlying design principles, offering a structured framework that surpasses the categorizations found in previous surveys. This novel approach allows for a more intuitive understanding of the field's evolution and current state.
  
  \item \textbf{Extensive Coverage of Application Cases}: Our survey extends the discussion to various application cases built upon car-following models, highlighting their significant impact on the environment, safety, mobility, and operation efficiency. This comprehensive review of applications is unprecedented in existing literature, providing valuable insights into the practical implications of car-following models.
  
  \item \textbf{Future Directions with Human-like Reasoning}: We underscore the future direction of integrating AI-driven methods with theory-based approaches to embed human-like reasoning processes into car-following models. This perspective is critical for developing more advanced, realistic models that can better mimic human driving behaviors, addressing a gap in current research trends.
\end{itemize}

By providing a thorough evaluation of current challenges and paving the way for future research directions in car-following models, our review marks a distinctive and valuable contribution to the literature.

\begin{table*}[!htb]
\centering
\caption{Summary of Car-Following Model Survey Papers}
\label{tab:car_following_review_papers}
\begin{tabular}{p{4cm}>{\centering\arraybackslash}p{1cm}p{5cm}p{5cm}}

\hline
\multicolumn{1}{c}{\textbf{Paper Title}} & \multicolumn{1}{c}{\textbf{Year}} & \multicolumn{1}{c}{\textbf{Scope of Work}} & \multicolumn{1}{c}{\textbf{Limitations}}\\
\hline
Car following models \cite{1992_Rothery} & 1992 & Summarized car-following models using mathematical model descriptions, stability analysis, calibration from empirical testing and evaluation and finally automatic vehicular control systems. & It only covers stimulus-response models from psychological and behavior perspective.\\
\hline

Car-following: a historical review \cite{1999_Brackstone_McDonald} & 1999 & Summarized theoretical early-stage car-following models from psychological and behavior perspective, such as safety distance, desired gaps, stimulus-response process. & It doesn't include some important models such as Intelligent Driver Model (IDM) and Optical Velocity Model (OVM).\\
\hline
Comparison of Car-following models \cite{2004_Olstam_Tapani} & 2004 & Compared theoretical car-following models used in well-known traffic micro-simulation software, including AIMSUN, MITSIM, Paramics and VISSIM. & From a software perspective to evaluate how different programs could replicate real world scenarios.\\
\hline
Driving Behavior: models and challenges \cite{2007_Toledo} & 2007 & Reviewed and expanded review lists in \cite{1999_Brackstone_McDonald} with IDM, OVM and Cellular Automata. Discussed the challenges associated with the myopic driver behavior models. & Only listed theory-based car-following models from human behavior perspective.\\
\hline
Car-following models  fifty years of linear stability analysis   a mathematical perspective \ \cite{2011_Wilson_Ward} & 2011 & A comprehensive survey of linear stability types, illustrated with straightforward examples and rigorous mathematical criteria. &  From a control theory perspective, it introduced a general framework that simplifies the analysis of linear stability by requiring only three partial derivatives.\\
\hline
Microscopic car-following model for the traffic flow: the state of the art \cite{2012_Li_Sun} & 2012 & Summarized and reviewed theory-based car-following models, also provided the stability analysis from control theory point of view. & Only reviewed theory-based car-following models from human behavior perspective.\\
\hline
Traffic flow dynamics\cite{2013_Treiber_Kesting} & 2013 & Categorized existing models as elementary models and strategic models by design principles in microscopic traffic flow modeling. & Only covers kinematic models from the physics perspectives.\\
\hline
Incorporating human-factors in car-following models: a review of recent developments and research needs\cite{2014_Saifuzzaman_Zheng} & 2014 & Introduced various human factors that could potentially impact driving styles. & Only summarized theory-based car-following models from engineering and human behavior perspective.\\
\hline
Genealogy of traffic flow models \cite{2015_vanWageningen-Kessels_VanLint_Vuik_Hoogendoorn} & 2015 & Summarized macro-, meso- and micro- traffic flow models, offered an historical model tree by tracing genealogy of the models. & Reviewed and surveyed car following models from the traffic engineering and physical perspectives\\
\hline
A state-of-the-art review of car-following models with particular considerations of heavy vehicles \cite{2015_Aghabayk_Sarvi_Young} & 2015 & Categorized CF models into classical models and AI models. Emphasized the impact of heavy trucks in car following behavior. & Under AI category, only fuzzy logic models and Neural Network models were considered. Under the classical model category, it doesn't include most recent ACC models\\
\hline
 Simulation Strategies for Mixed Traffic Conditions: A Review of Car-Following Models and Simulation Frameworks \cite{2020_Matcha_Namasivayam} & 2020 & This review paper focus on Heterogeneous vehicle types and mixed traffic flow modeling & Inherit the same categorization from previous review paper \cite{1999_Brackstone_McDonald} and mainly from simulation software perspective\\
\hline
A review of car-following models and modeling tools for human and autonomous-ready driving behaviors in micro-simulation \cite{2021_Ahmed_Huang_Lu} & 2021 & Emphasized the differences between human-driven vehicles and autonomous vehicles in car-following models. & Inherit the same categorization from previous review paper \cite{1999_Brackstone_McDonald} and mainly from simulation software perspective.\\
\hline
The Car-Following Model and Its Applications in the V2X Environment: A Historical Review \cite{2021_Han_Shi_Chen} & 2021 & Emphasize  applications of car-following models in the V2X environment & Use the similar categorization framework from previous review paper \cite{1999_Brackstone_McDonald, 2007_Toledo} to categorize car-following models\\

\hline
Modeling the Car-Following Behavior with Consideration of Driver, Vehicle, and Environment Factors: A Historical Review \cite{2022_Han_Wang_Wang} & 2022 & This review paper surveyed car models that consider diverse driver styles, physical and dynamic characteristics of vehicles, as well as environmental factors. & Inherit the similar categorization from previous review paper \cite{1999_Brackstone_McDonald, 2007_Toledo,2021_Han_Shi_Chen} and is developed from traffic engineering perspective. \\ 

\hline
Modelling and simulation of (connected) autonomous vehicles
longitudinal driving behavior: A state-of-the-art \cite{ 2023_Sadid_Antoniou} & 2023 & This paper reviewed parameters for (C)AVs modelling. It also summarizes the lately published data-driven models for (C)AVs CF behavior, and identification of KPIs used for impact assessments. & Mainly focus on CF models' parameters that are more crucial and sensitive in differentiating (C)AVs from human driven vehicles. \\
\hline
Microscopic Modelling of Car-Following Behavior: Developments and Future Directions \cite{2023_He_Zhou_Wang} & 2023 & This paper analyzes advantages  and  limitations of four main  types of CF models: Kinematics-based, dynamics-based, psychological-based, and learning-based. & Didn't provide subcategories and missed Cellular Automata, and ACC models. \\ 
\\
\hline
Car-following models for human-driven vehicles and autonomous vehicles: A systematic review \cite{2023_Wang_Shi_Tong_Gu_Cheng} & 2023 & Reviewed CF models in terms of accuracy, calibrations, stability analysis and specific model parameters for HDVs and AVs simulation &  Limited to Traffic Engineering perspective for categorization and applicable background\\

\hline
\end{tabular}
\end{table*}

\section{Theory-based Models}
Theory-based models are easy to interpret and highly flexible, which can be adapted to various applications. They are also easy to perform stability analysis with traditional control techniques. Kinematic Models emphasize the physical aspects of vehicle motion and dynamics; Psycho-Physical Models integrate human factors, particularly how drivers perceive surrounding environment and their decision-making processes; Adaptive Cruise Control (ACC) Models emphasize the mechanistic low-level control to maintain safe and efficient vehicle dynamics. 
\subsection{Kinematic Models} 
This principled car-following models are based on kinematic relationship, considering factors as velocity, headway and distance. First-order models are simpler models that prescribe the motion of a vehicle in terms of its velocity. Second-order models take acceleration into account, which can provide a more detailed simulation of real-world driving behavior. Despite their varied mathematical frameworks, many of these models share a common heuristic basis, making them interchangeable under specific scenarios. For instance, the OVM combined with Euler updates for position and speed is mathematically equivalent to Newell's model when Newell's reaction time parameter is set to the inverse sensitivity parameter ($a^{-1}$) of the OVM \cite{2013_Treiber_Kesting}. Moreover, Gipps' model and the IDM model share a set of parameters derived from the same driving strategy that drivers tend to keep a safe distance from the leading vehicle with comfortable speed or accelerations. Table \ref{tab:car_following_kinematic_models} provides a comparison among kinematic models. 


\subsubsection{First-Order Kinematic Models}\

The first-order kinematic car-following model, an early approach in car-following modeling, is a  mathematical description of the motion using the classical law of motion to update the vehicle's position at each time step. This type of model often features parameters that possess meaningful physical interpretations, and these principled models treat each driver-vehicle unit as a particle or automaton. As an elementary model, Herrey and Herrey \cite{1945_Herrey_Herrey} introduced the concept of influence space, considering minimum safety distance, and validated their hypothesis using time-space diagrams derived from field data. Pipes \cite{1953_Pipes} describes a linear relationship between the vehicle speed with respect to the minimum safe headway suggested by the California Motor Vehicle Code. Kometani and Sasaki performed stability analysis by introducing a transfer lag to Pipes' traffic dynamic model \cite{1958_Kometani_Sasaki} and built a safety index equation that is measured to represent the safety level in sinusoidal traffic flow\cite{1959_Kometani_Sasaki, 1959_Kometani_Sasaki_2}. 

\textbf{Gipps' Model}
Gipps \cite{1981_Gipps} developed a more advanced version of these explicit physics-based car-following models, assuming that a driving agent adapts the speed to smoothly and safely proceed behind its leader. Gipps model puts two constraints on the CF behavior: the vehicle will not exceed the driver's desired speed or a vehicle-specific free-flow speed, and the following driver selects a speed that can bring the follower vehicle to a safe stop. In addition, while the preceding vehicle decelerates, a safety margin should be added to the following driver's reaction time, limiting the speed of the following vehicle based on the performance of the leader vehicle, particularly with respect to its deceleration capability. The formulation is shown as follows:
\begin{equation}
\label{deqn_ex1a}
\resizebox{\linewidth}{!}{
$\begin{aligned}
v_{n}(t+\tau) = \min\Bigg\{ &v_{n}(t) + 2.5a_{n}\tau \left(1 - \frac{v_{n}(t)}{V_{n}}\right) \left(0.025 + \frac{v_{n}(t)}{V_{n}}\right)^{1/2}, \\
&b_{n}\tau + \sqrt{b_{n}^{2}\tau^{2} - b_{n}\left[2\left[x_{n-1}(t) - s_{n-1} - x_{n}(t)\right] - v_{n}(t)\tau - \frac{v_{n-1}(t)^{2}}{\hat{b}}\right]} \Bigg\}
\end{aligned}$}
\end{equation}

In the formulation, $a_n$ is the maximum acceleration; $b_n$ is the most severe braking that the driver of vehicle $n$ wishes to undertake; $v_n$ is the speed at which the driver of vehicle $n$ wishes to travel; $V_n$ is the desired speed of vehicle $n$; $\tau$ is the reaction time; $\hat{b}$ is the estimate of $b_{n-1}$. Because of its explicit physical meaning, modified Gipps' model was used in several traffic simulation software\cite{2007_Spyropoulou,2012_Ciuffo_Punzo_Montanino}. Krauss's model \cite{1998_Krauss} and Gipps' model are categorized within the same "family" of CF models, because both models fundamentally consider braking distances to derive interactions between vehicles. This shared focus on braking distances and the resultant dynamics establishes a conceptual link between the two models. Rakha and Wang \cite{2009_Rakha_Wang} developed an analytic calibration steps of Gipps model for steady state behavior. Some studies \cite{2001_Wilson} showed that the Gipps' model also allows reproducing the Phantom Traffic Jam, although the model is formulated as a first-order category. A modified Gipps model \cite{2023_Shah_Lee_Kim} addressed the potential collision when the most severe braking the following vehicle wishes to undertake is harder than the  perceived deceleration rate of lead vehicle. 

\textbf{Newell's Model}
In Newell's Model \cite{2002_Newell}, the time-space trajectory of the $n^{th}$ vehicle is essentially the same as the $(n-1)^{th}$ vehicle, except for a translation in space and in time. This is a very simple rule for car following behavior compared with other models; however, it is often much more accurate than many. Newell's Model is defined as:
\begin{equation}
\label{deqn_ex1a}
x_n\left(t+\tau_n\right)=x_{n-1}\left(t\right)-d_n
\end{equation}
where $x_{n}(t)$ is a piecewise linear trajectory of vehicle n, which will be a simple translation of piecewise linear $x_{n-1}(t)$ by a distance $d_n$ and a time lag $\tau_n$. By translating their trajectory forward or backward, drivers can manage the spacing between their vehicle and the preceding vehicle, ensuring a suitable and desirable headway. Newell's Model implies that drivers have the ability to adapt their position relative to the preceding vehicle by adjusting their velocity and position in a controlled manner, thereby maintaining a safe and comfortable driving experience. These fixed translations have been validated with video-taped trajectory data and described by a bivariate normal distribution \cite{2004_Ahn}. L-L model \cite{2010_Laval_Leclercq} introduced a  behavior deviation term to the basic Newell's model and find that timid and aggressive drivers are the cause of spontaneous traffic jam. Later on, an asymmetric behavioral model (ABM) is developed by adding five parameters per driver to capture different driver reactions to traffic oscillations \cite{2012_chen}. To account for randomness in driver behaviors and further understand traffic oscillations explained by string instability, stochastic components were added to the steady-state Newell's equations \cite{2014_Laval_Toth_Zhou, 2020_Xu_Laval}. Newell's model can replicate the formation and propagation of traffic waves with reasonable accuracy, and is particularly effective in large-scale network simulations \cite{2015_Przybyla}. By considering vehicle/engine dynamics and driving situation-dependent stochasticity, multidimensional stochastic Newell car-following model (MSNCM) is proposed for emission estimation \cite{2021_Meng}.   

\textbf{Cellular Automata}
Cellular automata models were first introduced in the study \cite{1983_Wolfram}, defined by a one-dimensional string of cells with two states (occupied and empty). The cell updating policy is known as Rule 184, described as: 
\begin{equation}
\label{deqn_ex1a}
v_\alpha\left(t+1\right) =
\begin{cases}
    1, & \text{if } x_{\alpha-1}\left(t\right) - x_\alpha\left(t\right) > 1 \\
    0, & \text{otherwise}
\end{cases}
\end{equation}
\begin{equation}
\label{deqn_ex1a}
x_\alpha\left(t+1\right)=x_\alpha\left(t\right)+v_\alpha(t+1)
\end{equation}
where $\alpha$ is the vehicle index. $x_\alpha$ is the location of the vehicle $\alpha$, $v_\alpha$ is the speed of vehicle $\alpha$.

A popular and simple Cellular Automata (CA) model is Nagel-Schreckenberg Model (NSM) \cite{1992_Nagel_Schreckenberg}. The NSM model can only update speed and acceleration on multiples of $7.5m/s$ and $7.5m/s^2$ due to the coarse discretization of space and time, and one cell corresponds to effective vehicle length and one-time step under car-following mode. 
First modification was made by adding a slow-to-start rule to the set of NaSch rules \cite{1993_Takayasu_Takayasu}
Some refined CA models include the Velocity Dependent Randomization (VDR) Model \cite{1998_Barlovic_Santen_Schadschneider_Schreckenberg}, Benjamin-Johnson-Hui (BJH) Model \cite{1996_Benjamin_Johnson_Hui} and Anticipation Model \cite{2000_Knospe_Santen_Schadschneider}, which generalized slow-to-start rule in Takayasu-Takayasu (TT) model. The KKW (Kerner–Klenov–Wolf) \cite{2002_Kerner_Klenov_Wolf} is a CA model within the framework of three-phase traffic theory and allows for smaller interval states. While the critical aspect of the KKW model allows the model to simulate the variability and unpredictability of real-world driving behavior, it can also lead to instability in the simulation. Bham and Benekohal (2004) \cite{2004_Bham_Benekohal} proposed the CELLSIM model to incorporate relative speed and distance headway into CA framework. Chakroborty and Maurya (2008) \cite{2008_Chakroborty_Maurya} conducted comparable analysis of different CA based traffic flow models in terms of microscopic properties (like time headway distribution, acceleration noise, stability, etc.). By focusing on spatio-temporal patterns, phase transitions, and growth patterns of oscillations, cellular automata can capture the intricate behaviors observed in real-world traffic, such as the formation and dissipation of traffic jams, and the transition from free-flow to congested traffic conditions \cite{2006_Daganzo, 2016_Tian_Li_Treiber_Jiang_Jia_Ma}. CA based simulation models are often used for heterogeneous traffic flow analysis, including pedestrian and bicycle flow dynamics after adjusting their parameters to reflect the lower speeds and smaller sizes \cite{2007_Meng_Dai_Dong_Zhang, 2018_Kong_List_Guo_Wu, 2009_Mallikarjuna_Rao,  2012_Vasic_Ruskin, 2018_Chen_Li_Wang_Jiang, 2019_Gavriilidou_Daamen_Yuan_Hoogendoorn, 2022_Tanveer_Kashmiri, 2023_Singh_Rao}. 

\begin{table*}[!htb]
\caption{Summary of Representative Theory Based Car Following Models from Kinematic Perspective}
\label{tab:car_following_kinematic_models}
\centering 
    
\begin{tabularx}
{\textwidth}{|>{\columncolor[HTML]{EFEFEF}\centering\arraybackslash}p{0.10\textwidth}|>{\centering\arraybackslash}p{0.2\textwidth}|X|X|}

\hline 
\rowcolor[HTML]{EFEFEF}
\textbf{Model} & \textbf{Authors \& Year} & \multicolumn{1}{c|}{\textbf{Key Strengths}} & \multicolumn{1}{c|}{\textbf{Key Limitations}} \\ \hline

Gipps Model & 
Gipps (1981) \cite{1981_Gipps} \newline 
Krauss (1998) \cite{1998_Krauss} \newline
Shah et al. (2023) \cite{2023_Shah_Lee_Kim} 
& 
1. Safety considerations: Gipps Model considers braking capabilities and adding a safety margin, even if the leading vehicle suddenly starts to slow down at this maximum rate. \newline 
2. Explicit Parametrization: Gipps Model describes various aspects of driver behavior and vehicle performance, including maximum acceleration, comfortable deceleration, desired speed, and reaction times. These parameters can be adjusted to simulate different driving styles, vehicle types, and traffic conditions. & 
1. Deceleration Assumption: The braking action a driver decides to take is based on ideal scenario that assumes the driver is always able to accurately judge the necessary braking effort. Different combinations of $b_n$ and  $\hat{b}$ could lead to potential collision \cite{2023_Shah_Lee_Kim} under extreme scenarios.
\\ \hline

Newell's Model & 
Newell (2002) \cite{2002_Newell}; \newline
Laval and Leclercq (2010) \cite{2010_Laval_Leclercq}; \newline
D. Chen et al. (2012) \cite{2012_chen}; \newline
Laval et al. (2014) \cite{2014_Laval_Toth_Zhou}; \newline
Xu and Laval (2020) \cite{2020_Xu_Laval}; \newline
Meng et al. (2021) \cite{2021_Meng}; \newline
& 
1. Simplicity: Newell’s Model is easy
to implement and analyze. 
\newline
2. Traffic wave simulation: Newell’s Model is often used to simulate traffic waves caused by traffic signals or other disturbances in the interrupted traffic flow.  & 

1. Limited Representations: Newell’s Model simplifies the driving process without any driver-related parameters, which may lead to less realistic representations.    \\ \hline

Cellular Automata & 
Wolfram (1992) \cite{1983_Wolfram} \newline
Takayasu and Takayasu (1993) \cite{1993_Takayasu_Takayasu} \newline
Benjamin et al. (1996) \cite{1996_Benjamin_Johnson_Hui} \newline
Barlovic et al. (1998) \cite{1998_Barlovic_Santen_Schadschneider_Schreckenberg} \newline
Knospe et al. (2000) \cite{2000_Knospe_Santen_Schadschneider} \newline
Kerner et al. (2002) \cite{2002_Kerner_Klenov_Wolf} \newline
Bham and Bebekohal (2004) \cite{2004_Bham_Benekohal} & 
1. Scalability: CA models are computationally efficient as they operate on a discrete grid with simple updating rules, making them suitable for large-scale traffic simulations. \newline
2. Flexibility: CA models are versatile in representing various traffic scenarios, such as heterogeneous traffic components and meta-stability. & 
1. Purely Particle-Based: CA models treat vehicles as particles moving on a grid without considering their actual physical dimensions beyond the grid size. This abstraction can overlook some aspects of vehicle behavior and interactions that are influenced by the physical size and the mechanics of vehicles, for instance, the distance headways.  \\ \hline

\cellcolor[HTML]{EFEFEF}GM Models (Gazis-Herman-Rothery) & 
Chandler et al. (1958) \cite{1958_Chandler_Herman_Montroll}
\newline
Herman et al. (1959) \cite{1959_Herman_Montroll_Potts_Rothery}
\newline
Gazis et al. (1959) \cite{1959_Gazis_Herman_Potts}
\newline
Gazis et al. (1961) \cite{1961_Gazis_Herman_Rothery} &
1. Stimulus-response foundation: The GM model provides generalizable analytic framework from the stimulus-response perspective. $Response=f(sensitivity,stimulus)$. It also allows customization based on different driver characteristics. &
1. Deterministic Stimulus: GM model assumes precise perception of stimulus, ignoring the inherent vagueness of CF process. \newline
2. Parameter Calibration: Calibrations and Validations of GM model's parameters are often contradictory, as CF behaviors vary with traffic and flow conditions. 
 \\ \hline
Optimal Velocity Model (OVM) &
Bando et al. (1995) \cite{1995_Bando_Hasebe_Nakayama_Shibata_Sugiyama} \newline
Nagatani (1998) \cite{1998_Nagatani} \newline
Helbing and Tilch (1998) \cite{1998_Helbing_Tilch} \newline
Konishi et al. (1999) \cite{1999_Konishi_Kokame_Hirata} \newline
Lenz et al. (1999) \cite{1999_Lenz_Wagner_Sollacher} \newline 
Nakayama et al. (2001) \cite{2001_Nakayama_Sugiyama_Hasebe} \newline
Jiang, Wu and Zhu (2001) \cite{2001_Jiang_Wu_Zhu} \newline
Hasebe et al. (2003) \cite{2003_Hasebe_Nakayama_Sugiyama} \newline
Gong et al. (2008) \cite{2008_Gong_Liu_Wang} \newline
Jin et al. (2010) \cite{2010_Jin_Wang_Tao_Li} \newline
Yu et al. (2019) \cite{2019_Yu_Jiang_Qu} \newline
Shang et al. (2023) \cite{2022_Shang_Rosenblad_Stern}
&
1. Interpretability: The OVM is straightforward with few parameters, aligning well with real-world driving behavior. \newline
2. Flexibility: The OVM can be easily extended or modified to include additional factors, e.g. macroscopic traffic conditions, safety constraints. & 
1. Simplifications: The assumption that drivers always pursue an optimal velocity as mental intention may not be true in all real-world situations. 
\\ \hline
Intelligent Driver Model (IDM) &
Treiber et al. (2000) \cite{2000_Treiber_Hennecke_Helbing} \newline
Treiber and Helbing (2003) \cite{2003_Treiber_Helbing} \newline
Treiber et al. (2006) \cite{2006_Treiber_Kesting_Helbing} \newline
Kesting, Treiber and Helbing (2010) \cite{2010_Kesting_Treiber_Helbing} \newline
Derbel et al. (2013) \cite{2013_Derbel_Peter_Zebiri_Mourllion_Basset} \newline
Eggert et al. (2015) \cite{2015_Eggert_Damerow_Klingelschmitt} \newline
Hoermann et al. (2017) \cite{2017_Hoermann_Stumper_Dietmayer} \newline
Treiber and Kesting (2017) \cite{2017_Treiber_Kesting} \newline
Hubmann et al. (2018) \cite{2018_Hubmann_Schulz_Xu_Althoff_Stiller} \newline
Zhou et al. (2024) \cite{2024_Zhu_Li_Qu_Ran}
&
1. Smoothness and safety: IDM captures key strategies of driving behaviors, such as maintaining a safe following distance, smooth acceleration, and comfortable deceleration.  \newline
2. Integral and Versatility: The IDM integrates the follower’s speed, relative velocity, and distance headway, offering the versatility to apply in both uninterrupted and interrupted flows across highways and urban settings.
&
1. Coefficient’s Meaning: Certain parameters (e.g., Acceleration Exponent $ \delta $) don’t contain physical meaning. 
\\ \hline

\end{tabularx}
\end{table*} 

\subsubsection{Second-Order Kinematic Models}\

Second-order kinematic models are based on the concept that the following vehicle adjusts acceleration (or deceleration) in response to the observational motion of the leading vehicle to maintain a desired speed and safety distance. The second-order kinematic models often incorporate sensitivity factors or driver preference parameters to capture the driver's response to various stimuli or conditions in the traffic environment. These sensitivity factors represent the driver's characteristics, driving styles, and preferences, which are widely used for analyzing factors that might affect the car-following process by formulating in different ways, such as adding stochastic factors from additional stimuli and variations in sensitivity parameters.

\textbf{General Motors Model}
General Motors (GM) model family is also known as GHR (Gazis-Herman-Rothery) model, which was first developed by Chandler et al. \cite{1958_Chandler_Herman_Montroll} and Herman et al. \cite{1959_Herman_Montroll_Potts_Rothery} at the General Motors research laboratory. GM model family is based on the stimulus-response process of humans, and they have evolved through five generations. The stimuli term can cause the response to be acceleration, deceleration or constant speed. The model formulation is shown as follows:
\begin{equation}
\label{deqn_ex1a}
\frac{dv_n(t+\mathrm{\Delta})}{dt}=\lambda[v_{n-1}(t)-v_n(t)]
\end{equation}
$\lambda$ denotes the sensitivity of the driver, $\mathrm{\Delta}$ denotes the response time of the driver. Gazis et al. \cite{1959_Gazis_Herman_Potts, 1961_Gazis_Herman_Rothery} found that the sensitivity $\lambda$ was inversely proportional to the space headway. It was found that different data and different experiments offered a different combination of $m$ and $l$ with the same car following model \cite{1967_May_Keller,1972_Heyes_Ashworth,1976_Ceder_May}, which blocks the application of the GM models. Variations of GM Model are employed to analyze driver behaviors, for instance, a "multi-following model" \cite{1968_Bexelius} where drivers consider several vehicles ahead. Researchers applied GM models to emphasize the asymmetric nature of car following behavior for distinct treatment of acceleration and deceleration phases \cite{1974_Treiterer_Myers}.

\textbf{Optimal Velocity Model}
Optimal Velocity Models (OVMs) are a set of models based on the assumption that each vehicle pursues a legal velocity. Bando et al. \cite{1995_Bando_Hasebe_Nakayama_Shibata_Sugiyama, 1998_Bando_Hasebe_Nakanishi_Nakayama} proposed the first Optimal Velocity (OV) Model that each driver controls acceleration or deceleration to achieve an optimal velocity $V(\Delta x_n)$ according to headway distance $\Delta x$. The original formulation is expressed as follows: 
\begin{equation}
\label{deqn_ex1a}
\frac{dv_n}{dt}=a[V\left(\Delta x_n(t)\right)-v_n(t)]
\end{equation}
The optimal velocity function is as below:
\begin{equation}
\label{deqn_ex1a}
V\left(\Delta x_n\left(t\ \right)\right)=\frac{v_{max}\left[\tanh{\left(\Delta x_n\left(t\right)-\ h_c\right)}+\tanh{\left(h_c\right)}\right]}{2}
\end{equation}
where $a$ is the sensitivity coefficient, $v_{max}$ is the maximum velocity. $h_c$ is the safety distance. $\Delta x_n(t)$ is the relative distance to the preceding vehicle, and $v_n(t)$ is the velocity of the following vehicle.

OVM has lead to many variations, demonstrating its interpretability and adaptability to different aspects of traffic flow and vehicle dynamics. For instances, General Force Model \cite{1998_Helbing_Tilch} introduced terms that represent the attractive force and repulsive force; Full Velocity Difference Model \cite{2001_Jiang_Wu_Zhu} emphasizes the relative velocity between a following vehicle and the vehicle directly in front of it; Coupled Map CF model \cite{1999_Konishi_Kokame_Hirata} is built on OVM to describe the interactions between vehicles with the concept of coupled map lattices. By considering Multi-Vehicle interaction, researchers expanded OVM to consider both preceding or following vehicles \cite{1999_Lenz_Wagner_Sollacher, 2001_Nakayama_Sugiyama_Hasebe, 2003_Hasebe_Nakayama_Sugiyama}. By incorporating macroscopic and microscopic models, Nagatani \cite{1998_Nagatani} proposed a lattice hydrodynamic model combining macroscopic hydrodynamics with microscopic optimal velocity; Chiarello et al. \cite{2021_Chiarello_Piccoli_Tosin} used OVM to examine the follower-the-leader interactions in the transition from the inhomogeneous Aw–Rascle–Zhang (ARZ) model to the Lighthill–Whitham–Richards (LWR) model; By considering the asymetric and lane separation, Gong et al. \cite{2008_Gong_Liu_Wang} developed an asymmetric full-velocity difference, car-following model. Jin et al. \cite{2010_Jin_Wang_Tao_Li} considered the lateral separation characteristics between the followers and leaders on a single-lane highway based on the full optimal velocity model. A confined Full Velocity Difference (c-FVD) model \cite{2019_Yu_Jiang_Qu} was proposed that limits the acceleration or deceleration generated by the existing FVD models to a reasonable level. Each variation or extension addresses specific challenges or aspects of real-world driving, contributing to the broader understanding of traffic behavior. Shang et al. (2022) proposed an asymmetric optimal velocity relative velocity model (AOVRV), considering common velocity variation \cite{2022_Shang_Rosenblad_Stern}.

\textbf{Intelligent Driver Model (IDM)}
The Intelligent Driver Model (IDM) was proposed by Treiber et al. \cite{2000_Treiber_Hennecke_Helbing}, which is also known as a strategy-based model with a list of basic assumptions. For IDM models, the vehicle's acceleration is described as below:
\begin{equation}
\label{deqn_ex1a}
\frac{d}{dt}v=a_{max}\left(1-\left(\frac{v\left(t\right)}{v_{desired}}\right)^{\delta}-\left(\frac{s^\ast\left(v,\Delta v\right)}{s}\right)^2\right)
\end{equation}
where $a_{max}$ is the maximum acceleration of the vehicle. $v$ is the current velocity of the vehicle. $v_{desired}$ is the desired velocity in free traffic. $\delta$ is the acceleration exponent (usually set to 4). $s$ is the current bumper-to-bumper gap. $\Delta v$ is the relative velocity (velocity difference) with respect to the vehicle ahead.

The desired minimum gap $s^\ast$ is given by: 
\begin{equation}
\label{deqn_ex1a}
s^\ast\left(v,\Delta v\right)= s_0+max\left[0, Tv+ \frac{v\Delta v}{2\sqrt a_{max} b}\right]
\end{equation}
where, $s_0$ is the minimum desired distance (jam distance). $T$ is the desired time headway (the time a driver wants to be behind the leading vehicle). $b$ are the comfortable braking deceleration.

By introducing an internal dynamical variable to represent the subjective level of service, Treiber and Helbing\cite{2003_Treiber_Helbing} build the IDMM (intelligent driver model with memory) to incorporate memory effects in microscopic traffic models. The Multi-anticipative IDM models \cite{2006_Treiber_Kesting_Helbing} consider essential finite reaction times, estimation errors, and spatial and temporal anticipation. Kesting et al. \cite{2010_Kesting_Treiber_Helbing} enhanced the IDM model to assess the impact of adaptive cruise control vehicles on traffic capacity. Liebner et al. \cite{2012_Liebner_Baumann_Klanner_Stiller} applied IDM for ADAS to incorporate the spatially varying velocity profile to represent both car-following and turning behavior. Derbel et al. \cite{2013_Derbel_Peter_Zebiri_Mourllion_Basset} modified the IDM to improve driver safety and respect vehicle capability in emergent situations by adding a new term $cv^2/b$ to the expected safe time distance formula. Li et al. \cite{2015_Li_Li_Xu_Qian} modified IDM with power cooperation to strengthen the power of each vehicle in proportion to the immediately preceding vehicle. To build a realistic agent model, Eggert et al. \cite{2015_Eggert_Damerow_Klingelschmitt} proposed the Foresighted Driver Model (FDM), which assumes that a driver acts in a way that balances predictive risk with utility. Milanés \cite{2014_Milanés_Shladover} calibrated CACC model and tested IDM on actual vehicles as reference for modeling ACC car-following behavior in traffic flow simulations. Hoermann et al. \cite{2017_Hoermann_Stumper_Dietmayer} extended the IDM for autonomous vehicles to describe a probabilistic motion prediction applicable to long-term trajectory planning. Treiber and Kesting \cite{2017_Treiber_Kesting} added external noise and action points to the IDM to study mechanisms behind traffic flow instabilities and indifferent regions of finite human perception thresholds. Zhou, Qu and Jin (2017) \cite{2017_Zhou_Qu_Jin} added multiplication factors to build Cooperative Intelligent Driver Model(CIDM) to simulate AVs behavior to show that AVs could relieve traffic oscillation by give larger gaps for on-ramp vehicles. Hubmann et al. \cite{2018_Hubmann_Schulz_Xu_Althoff_Stiller} extend the IDM model for complex urban scenarios by incorporating a behavior prediction framework with context-dependent upper and lower bounds on acceleration. A recent study \cite{2022_Albeaik_Bayen_Chiri_Gong_Hayat_Kardous_Keimer_McQuade_Piccoli_You} analyzed the limitations of the IDM model from the numerical analysis: the velocities of specific vehicles might become negative at specific times and might diverge to $-\infty$ in a finite time. To capture the stochastic nature of human driver behavior,  IDM with time-varying parameters is proposed, with each parameter as a stochastic process and calibrating them in a time-varying manner \cite{2022_Zhang_Chen_Wang_Zheng_Wu}. The Self-Adaptive IDM (SA-IDM) \cite{2024_Zhu_Li_Qu_Ran}  dynamically adjusts acceleration in response to the leading vehicle's speed and time headway, ensuring stable and overdamped platoon behavior under all conditions.

\subsubsection{Psycho-physical Models}\

Psycho-physical models are developed to approximate drivers' reasoning and decision process, considering the uncertainty and boundary effects in recognizing relative speeds and distances. Forbes \cite{1968_Forbes_Simpson} first introduced the human factor in car-following behavior and found that time headway should always be equal to or greater than perception-reaction time. Table \ref{tab:Psycho_physical_car_following_models} summarize two representative models: Action Point Model and Fuzzy Logic Model. 

\begin{table*}[!htb]
\caption{Summary of Representative Theory Based Models from Psycho-Physical Perspective}
\label{tab:Psycho_physical_car_following_models}
\centering 
    
\begin{tabularx}
{\textwidth}{|>{\columncolor[HTML]{EFEFEF}\centering\arraybackslash}p{0.10\textwidth}|>{\centering\arraybackslash}p{0.2\textwidth}|X|X|}

\hline 
\rowcolor[HTML]{EFEFEF}
\textbf{Model} & \textbf{Authors \& Year} & \multicolumn{1}{c|}{\textbf{Key Strengths}} & \multicolumn{1}{c|}{\textbf{Key Limitations}} \\ \hline

\textbf{Action Points (AP)} &
Michaels and Cozen (1963) \cite{1963_Michaels_Cozan} \newline
Evans and Rothery (1973) \cite{1973_Evans_Rothery} \newline
Wiedemann and Reiter (1992) \cite{1992_Wiedemann_Reiter} \newline
Fritzsche and Ag (1994) \cite{1994_Fritzsche_Ag} \newline
Winsum (1999) \cite{1988_Ferrari} \newline
Wang et al. (2004) \cite{2004_Wang_Zhang_Li_Hirahara_Ikeuchi} \newline
&
1. Human Perception-Reaction: Models can be more realistic compared to models based purely on physical or mathematical principles. 
\newline
2. Adaptability: Action Point model can be adapted to represent different driver behaviors, which is particularly effective in simulating a variety of traffic components. 

& 
1. Complexity: Due to modeling multiple action points, it can be challenging for calibration. 
\newline
2. Discontinuity: Discrete nature of action points can lead to abrupt changes. 
\\ \hline

\textbf{Fuzzy Logic Models} &
Kikuchi and Chkkroborty (1992) \cite{1992_Kikuchi_Chakroborty} \newline
Chakroborty and Kikuchi (1999) \cite{1999_Chakroborty_Kikuchi} \newline
Zheng and McDonald (2005) \cite{2005_Zheng_McDonald} \newline
Chai and Wong (2015) \cite{2015_Chai_Wong} \newline
Hao, et al. (2016) \cite{2016_Hao_Ma_Xu} \newline
Bennajeh, et al. (2018) \cite{2018_Bennajeh_Bechikh_Said_Aknine}
&
1. Human-like reasoning: Fuzzy logic can incorporate multiple inputs and outputs, allowing for various factors affecting driver behavior
\newline
2. Flexibility: Fuzzy logic is good at dealing with uncertainty and imprecision in human driving behavior. 

& 
1. Subjectivity: Creating an effective set of fuzzy rules and membership functions can be subjective. 
\newline
2. Calibration and Validation Challenges: These models require extensive calibration and validation to ensure that the fuzzy rules accurately represent real-world driving behavior.
\\ \hline
\end{tabularx}
\end{table*}

\textbf{Action Point Model} The Action Point Model (APM) is a variant of Signal Detection Theory (SDT) specifically applied to the context of car-following models in traffic psychology.
The APM utilizes the principles of SDT to model the driver's perceptual sensitivity (signal detection) and response bias (decision criterion) in determining when to initiate an acceleration/braking action. By incorporating these components, the model provides a framework to understand and predict car-following behavior in terms of the driver's perceptual and decision-making processes. The prototype of APM research was presented by \cite{1963_Michaels_Cozan}, where the first perceiving threshold that distinguished the boundary of whether the driver could perceive the velocity changing relative to the vehicle ahead was proposed. This threshold was given as the visual angle to the rear width of the vehicle ahead. The visual angel ($\theta$) and its rate of change or angular velocity are calculated with the equations:
\begin{equation}
\label{deqn_ex1a}
\theta=2\tan^{-1}{(\frac{w}{2H})}
\end{equation}
\begin{equation}
\label{deqn_ex1a}
\frac{d\theta}{dt}=-w\ast\frac{Vr}{H^2}
\end{equation}
where $H$ is the gap between leading and following vehicles. $Vr$ is the relative speed. $w$ is the width of the leading vehicle. 

Todosiev \cite{1963_Todosiev} adopted a fundamental psycho-physical approach and investigated the thresholds where drivers change their behavior at an action point. Evans and Rothery \cite{1973_Evans_Rothery} conducted experiments to quantify the thresholds between different phases. Wiedemann and Reiter \cite{1992_Wiedemann_Reiter}  developed the well-known psycho-physical car-following model used in VISSIM. Another simulation program, PARAMICS, has incorporated the Fritzsche car-following model \cite{1994_Fritzsche_Ag}, an acceleration-based model from the psycho-physical car-following framework. As described in literature \cite{1988_Ferrari}, once the absolute value of angular velocity exceeds its threshold, a driver notices that ego-vehicle speed is different from that of the leading vehicle and reacts with an ac/deceleration opposite in sign to that of the rate of change of visual angle. Winsum \cite{1999_Winsum} integrated preferred time headway and Time-to-Collision (TTC) in a mathematical car-following model based on psychological evidence that human drivers regulate available time as a control mechanism. Wang et al. \cite{2004_Wang_Zhang_Li_Hirahara_Ikeuchi} devised ac/deceleration model based on the driver's cognitive mechanism using the Just-Noticeable Distance (JND) concept. Lochrane et al. \cite{2015_Lochrane_Al-Deek_Dailey_Krause} developed a multidimensional framework for modeling thresholds to account for the different behavior in the work and nonwork zones. Durrani and Lee \cite{2019_Durrani_Lee} calibrated the eight Calibration Constants (CCs) built into the Wiedemann model used by VISSIM to quantify the threshold values for drivers' APs. Wagner et al. \cite{2020_Wagner_Erdmann_Nippold} analyzed a large dataset of car-following pairs to investigate the distribution of APs in terms of speed differences and gaps with SUMO’s implementation of the Wiedemann model, finding that small distances indicate a slightly higher probability of finding an AP. 

\textbf{Fuzzy Logic Model}  
To model human decision-making process, fuzzy logic allows the modeling of the estimates of a driver with vagueness, uncertainty, and subjectivity using fuzzy rules. Several studies \cite{1992_Kikuchi_Chakroborty,1999_Chakroborty_Kikuchi,2005_Zheng_McDonald} integrated the fuzzy theory into the GM model and attempted to "fuzzify" the variables of the traditional GM model. A fuzzy cellular Automata Model\cite{2015_Chai_Wong} is developed to simulate the driver's decision-making process and estimate the effect of driving behavior on traffic performance, which utilizes fuzzy interface
systems (FISs) and membership functions to simulate the perception-decision process of individual drivers. Hao et al. \cite{2016_Hao_Ma_Xu} incorporated fuzzy logic into the five-layer structure: Perception–Anticipation–Inference–Strategy–Action to build the CF model. Bennajeh et al. \cite{2018_Bennajeh_Bechikh_Said_Aknine} employed four strategic variables, including gap distances, the velocity of followers and leaders, and the action of followers, to build the fuzzified anticipation car-following model. Although the fuzzy-logic model is a reasonable solution to consider the psycho-physical process, the sensitivity of membership values greatly impacts the accuracy and constructs of parameters relying on engineering preferences. 

\subsection{Adaptive Cruise Control Model}
Unlike the previous models designed to simulate human driver behavior and reveal the disturbances of natural traffic, ACC models are developed to dampen the instability from the mechanistic view. ACC systems directly manage the basic operations of the vehicle, such as throttle control, braking, and steering. Low-level controllers are responsible for executing the commands to meet the higher-level objectives, like maintaining a set speed or following distance in ACC. Adaptive cruise control has been researched since the 1970s\cite{1976_Hoberock_Rouse,1977_Chiu_Stupp_Brown,1978_Shladover,1989_Cho_Hedrick,1990_McMahon_Hedrick_Shladover,2003_Suzuki_Nakatsuji} and was first on the market when Mitsubishi launched the "Preview Distance Control" driver assistance system in 1995. ACC and CACC (Cooperative Adaptive Cruise Control) are heavily researched CAV applications. ACC models are designed for autonomous car following to reduce the feasible time headway among vehicles \cite{2011_Ploeg_Serrarens_Heijenk}.
This paper \cite{2013_Milanés_Shladover_Spring_Nowakowski_Kawazoe_Nakamura} examines challenges of CACC systems in real-world deployment, including routing protocols, driver engagement, and adapting control designs to realistic traffic conditions. Some studies \cite{2015_Shladover_Nowakowski_Lu_Ferlis, 1994_Swaroop_Hedrick_Chien_Ioannou,2002_Li_Shrivastava} suggest that the Constant Time Gap policy is more robust against error propagation through traffic than the Constant Distance Gap policy. By taking into account underlying the vehicle dynamics, e.g., force, torque, and power, Rakha et al. \cite{2004_Rakha_et_al, 2012_Rakha_et_al} combined vehicle powertrain model with driver behavior to realize more realistic modeling of intelligent vehicle applications. Fadhloun-Rakha (FR) model \cite{2012_Fadhloun_Rakha, 2020_Fadhloun_Rakha} was proposed by capturing heterogeneous driving behavior and vehicle engine dynamics, considering two constraints regarding car following interactions (collision avoidance and steady-state  traffic  stream).  Table \ref{tab:ACC_car_following_models} summarizes representative adaptive cruise control models.
\subsubsection{Linear Model}
Linear controller assumes acceleration is proportional to the deviation from target spacing and relative speed. Helly \cite{1959_Helly} proposed the first linear model that prescribes the adaptation of the acceleration of each vehicle as a function of relative velocity, relative distance from the leading car, and other pertinent parameters. Milanés et al.\cite{2012_Milanés_Villagrá_Godoy_González} compared four control techniques: PI, intelligent PI, fuzzy controller, and adaptive-network-based fuzzy control, showing that the intelligent PI performs best. Dias et al.\cite{2014_Dias_Pereira_Palhares} used a linear longitudinal force model as a feedforward controller or Proportional-Integral-Derivative (PID) to compensate for the powertrain non-linearity. The LQ-based controller has been applied on ACC and CACC systems and tested under rural and urban roads in a CV environment, considering both communication delay and driver reaction time \cite{2012_Kim_Tomizuka_Cheng,2016_Jin_Orosz}. 
A classical Constant-Time Headway Relative-Velocity (CTH-RV) model \cite{2023_Wang_Yanbing} with gains of speed error and spacing error is described as follows. 
\begin{equation}
\label{deqn_ex1a}
a_i\left(t+\tau\right)=k_1\left(v_{i-1}\left(t\right)-v_i\left(t\right)\right)+k_2\left(r\left(t\right)-r_d(t)\right)
\end{equation}
\begin{equation}
\label{deqn_ex1a}
r_d(t) = t_g * v_i(t)
\end{equation}
\begin{equation}
\label{deqn_ex1a}
a_i\left(t\right)\in[d_{max},\ a_{max}]
\end{equation}
where $k_1>0$ and $k_2>0$ are the gains of speed error and spacing error; $v_{i-1}$ and $v_i$ is the speed of the Preceding Vehicle (PV) and Following Vehicle (FV); $r(t)$ is the current distance between vehicles; $r_d(t)$ is the desired distance between vehicles;  $t_g$ is the desired time gap. $a_{max}$ is the maximum allowed acceleration; $d_{max}$ is the maximum allowed deceleration. The acceleration and deceleration values are bounded by comfortable values at $2m/s^2$ and $-3m/s^2$.

By considering the V2V "hop-by-hop" communication, an effective CACC controller \cite{2001_VanderWerf_Shladover_Kourjanskaia_Miller_Krishnan} received information about the velocity of PV, acceleration of PV, and braking capability of PV, which is defined as follows: 
\begin{equation}
\label{deqn_ex1a}
\resizebox{\linewidth}{!}{
$\begin{aligned}
a_{i}(t+\tau)= k_{0}\times a_{i-1}(t)+ k_1\times(v_{i-1}(t)-v_i(t)) + \\k_2 \times (r(t)-r_d(t))
\end{aligned}$}
\end{equation}
\begin{equation}
\label{deqn_ex1a}
a_i(t)\in[d_i, a_{max}]
\end{equation}
where $d_i$ is the braking capability of FV. The desired spacing $r_d$ is defined by the maximum safe following distance $r_{safe}(t)$, following distance with $0.5s$ time gap $r_{0.5s}(t)$, and a minimum allowed distance $r_{min}$ that is chosen to be 2m. 
\begin{equation}
\label{deqn_ex1a}
r_d(t) = max(r_{safe}(t),r_{0.5s}(t),r_{min}) 
\end{equation}
\begin{equation}
\label{deqn_ex1a}
r_{safe}(t)=\frac{v_{i-1}^2 (t)}{2} (\frac{1}{b_{i}}-\frac{1}{b_{i-1}})+ \delta \times v_{i-1}^2(t)   
\end{equation}
\begin{equation}
\label{deqn_ex1a}
r_{0.5s}(t)= 0.5*v_i(t)   
\end{equation}
\begin{equation}
\label{deqn_ex1a}
s_{min}=2  
\end{equation}
where, $r_d$, $r_{safe}$, and $r_{[t]s}$ are desired, safety, and spacing with $t$ seconds gap, $\delta=20$ ms is communication delay, $b_{i-1}$ is the braking capability of the PV, $b_i$ is the braking capability of the FV. The time gap ranges from $[0.5s,1.4s]$. $1.4s$ is used when the preceding vehicle is not CACC-controlled. 


\subsubsection{Nonlinear Model}
A Fuzzy ACC controller \cite{1995_Holve_Protzel_Bernasch_Naab}, whose output variables can be applied to low level controllers directly, was developed to ensure that the vehicle not only responds effectively to dynamic driving conditions but also respects and adapts to the driver's preferences and inputs, maintaining a balance between automation and human control. Naranjo et al.\cite{2003_Naranjo_González_Reviejo_García_DePedro} developed an adaptive fuzzy controller for speed and distance vehicle control, comprised of three layers: Mechanical Layer, Electronic Layer and Control Layer. Pérez et al.\cite{2013_Pérez_Milanés_Godoy_Villagra_Onieva} developed a CACC model through fuzzy controllers to perform platooning in highway scenarios. Zhang and Orosz \cite{2016_Zhang_Orosz} developed a distributed nonlinear controller for cooperation in multi-agent networks considering consensus and disturbance attenuation. To address the difficulties of obtaining parameters of the detailed longitudinal car-following model and account for nonlinear effects, Azızıaghdam and Alankuş \cite{2022_Azizıaghdam_Alankuş} designed a two-loop controller that only uses parameters known to OEMs (Original Equipment Manufacturer). The outputs of the Inner loop are the accelerator pedal position sent to the vehicle CAN-Bus. The outer loop controls the relative distance and the relative velocity. A personalized adaptive cruise control (PACC) systems \cite{2022_Wang_Wang_Han_Tiwari_Work} use a Gaussian Process (GP) model to learn individual driving styles by implicitly considering
the delay of driver reaction and the low-level vehicle dynamics. It significantly reduces human overrides in simulations, enhancing driving comfort and user experience compared to standard ACC models.

\subsubsection{Model Predictive Control (MPC)}
This type of control algorithm is flexible and usually defined by multiple objective functions or cost functions that can directly incorporate key constraints. According to the spacing-control laws, Bageshwar et al. \cite{2004_Bageshwar_Garrard_Rajamani} applied the MPC in a vehicle cut-in scenario for transitional maneuvers of ACC vehicles. Based on radar measurements and V2X communication, Moser et al. \cite{2015_Moser_Waschl_Kirchsteiger_Schmied_DelRe} developed a stochastic MPC with a linear cost function using a piecewise linear approximation of the fuel consumption map. Cheng et al.\cite{2019_Cheng_Li_Mei_Nie_Zhao} developed a multiple-objective MPC model integrated with direct yaw moment control (DYC) to ensure vehicle dynamic stability and improve driving comfort. A generic control framework proposed by \cite{2012_Hoogendoorn_Hoogendoorn_Wang_Daamen,2013_Wang_Treiber_Daamen_Hoogendoorn_vanArem} assumes that accelerations of cruising vehicles are controlled to optimize a cost function reflecting different control objectives, i.e., safety, comfort, efficiency, and sustainability. Given that the control input $\mathbf{u}=\{ u_i(\tau)|\tau\geq t_k \}$ is applied, the cost function of car-following is formulated as:
\begin{equation}
\label{deqn_ex1a}
\begin{aligned}
\begin{split}
J\left(t_k,\mathbf{x}(t_k)\middle|\mathbf{u}\right) &= \int_{t_k}^{t_k+T_p}{e^{-\eta\tau}\mathcal{L}(\mathbf{x}(t),\mathbf{u}(t),\tau)}d\tau \\
&\quad + e^{-\eta\left(t_k+T_p\right)}\phi(\mathbf{x}(t_k+T_p))
\end{split}
\end{aligned}
\end{equation}
The notation $J\left(t_k,\mathbf{x}\middle|\mathbf{u}\right)$ represents the expected cost associated with the control input $\mathbf{u}$ when the system states evolve from $\mathbf{x}(t_k)$ to $\mathbf{x}(t_k+T_p)$. Here, $T_p$ denotes the prediction horizon. The function $\mathcal{L}(\cdot)$ represents the running cost during the interval $[\tau, \tau+d\tau)$, while $\phi(\cdot)$ represents the terminal cost, capturing the remaining cost at the terminal time. The parameter $\eta\in [0, +\infty)$ serves as a discounting factor with units of $s^{-1}$, balancing the current and future costs. Rolling horizon stochastic optimal control strategy for both ACC and CACC were implemented in a model predictive control fashion, considering the Uncertainties in system dynamics and sensor measurements \cite{2017_Zhou_Ahn_Chitturi_Noyce}. A serial distributed model predictive control (MPC) \cite{2019_Zhou_Wang_Ahn} is developed for connected automated vehicles (CAVs), ensuring local stability and multi-criteria string stability by formulating future state constraints and tuning weight matrices, with mathematical proofs and simulations demonstrating its superiority over traditional MPC methods in maintaining stability.

\begin{table*}[!htb]
\caption{ADAPTIVE CRUISE CONTROL CAR-FOLLOWING MODELS}
\label{tab:ACC_car_following_models}
\centering 

\begin{tabularx}
{\textwidth}{|>{\columncolor[HTML]{EFEFEF}\centering\arraybackslash}p{0.10\textwidth}|>{\centering\arraybackslash}p{0.18\textwidth}|X|X|}

\hline 
\rowcolor[HTML]{EFEFEF}
\textbf{Model} & \textbf{Authors \& Year} & \multicolumn{1}{c|}{\textbf{Key Strengths}} & \multicolumn{1}{c|}{\textbf{Key Limitations}} \\ \hline

\cellcolor[HTML]{EFEFEF}Linear Model &
Helly (1959) \cite{1959_Helly} \newline
Milanes et al. (2012) \cite{2012_Milanés_Villagrá_Godoy_González} \newline
Kim et al. (2012) \cite{2012_Kim_Tomizuka_Cheng} \newline
Dias et al. (2014) \cite{2014_Dias_Pereira_Palhares} \newline
Jin and Orosz (2016) \cite{2016_Jin_Orosz} 
&
1. Flexibility: The linear control model can be designed to represent various traffic conditions by adjusting the gain values for different control inputs.
\newline
2. Robustness: The combination of feedback and feed-forward control mechanisms improves the model’s robustness to uncertainties and disturbances in the environment.
&
1. Limitation of linearity: The model assumes a linear relationship between input variables (e.g., relative distance, speed, acceleration) and control actions (e.g., desired acceleration). With a time-invariant feedback gain, the control action does not adapt to changing conditions in the system’s environment. 
\newline
2. Safety Risks: Linear control systems often do not explicitly incorporate collision-free constraints, which are critical in ensuring safety in traffic management and automated driving systems.\\
\hline
Nonlinear Model &
Holve, et al. (1995) \cite{1995_Holve_Protzel_Bernasch_Naab} \newline
Naranjo, et al. (2003) \cite{2003_Naranjo_González_Reviejo_García_DePedro} \newline
Perez, et al. (2013) \cite{2013_Pérez_Milanés_Godoy_Villagra_Onieva} \newline
Zhang and Orosz (2016) \cite{2016_Zhang_Orosz} \newline
Aziziaghdam and Alankus (2022) \cite{2022_Azizıaghdam_Alankuş} \newline
Wang, et al. (2022) \cite{2022_Wang_Wang_Han_Tiwari_Work}
&
1. Robustness: Nonlinear control techniques offer increased robustness to uncertainties and disturbances in the environment, such as sensor noise, communication delays, or changing road conditions.
\newline 
2. Stability and performance guarantees: Some nonlinear control techniques provide stability and performance guarantees for the closed-loop system, which are crucial in safety-critical applications. &
1. Design Complexity: Nonlinear models can be more complex, requiring extensive calibration, validation, and tuning efforts. The nonlinear relationship between inputs and outputs makes it harder to understand the underlying driving behavior or control actions.  \\
\hline
Model Predictive Control &
Bageshewar, et al. (2004) \cite{2004_Bageshwar_Garrard_Rajamani} \newline
Hoogendoorn, et al. (2012) \cite{2012_Hoogendoorn_Hoogendoorn_Wang_Daamen} \newline
Wang, et al. (2013) \cite{2013_Wang_Treiber_Daamen_Hoogendoorn_vanArem} \newline
Moser, et al. (2015) \cite{2015_Moser_Waschl_Kirchsteiger_Schmied_DelRe} \newline
Zhou et al. (2017) \cite{2017_Zhou_Ahn_Chitturi_Noyce}
Cheng, et al. (2019) \cite{2019_Cheng_Li_Mei_Nie_Zhao} \newline

&
1. Optimal control: MPC provides an optimal control solution by minimizing a cost function that can incorporate various objectives, such as maintaining a safe distance, reducing fuel consumption, or improving driving comfort. 
\newline
2. Explicit constraints handling: MPC can handle constraints explicitly, such as acceleration limits, speed limits, or safe following distances. 
\newline
3. Adaptability: MPC-based models can be adapted to different driver behaviors or customized driving styles by adjusting the cost function or model parameters.
&
1. Computational demands: The optimization problem in MPC needs to be solved in real-time, which can be computationally intensive, especially for large prediction horizons or high-dimensional models. 
\newline
2. Prediction model accuracy: The performance of MPC models highly depends on the accuracy of the prediction model used to predict the future states of the leading vehicle and the following vehicle. 
\\
\hline
\end{tabularx}
\end{table*}

\section{Learning-based Models}
Instead of an analytic modeling framework, more recently, data-driven car-following models have achieved human-level or beyond human-level performances under complex driving environments. It is unattainable to devise a parametric model considering all heterogeneous traffic compositions, road conditions, and varying data inputs. Whereas, data-driven approaches excel in scenarios with heterogeneous traffic compositions, diverse road conditions, and varied data inputs, not only in standard lane-based traffic situations but also in more challenging lane-free environments and in interactions with pedestrians. Table \ref{tab:data_driven_car_following_models} summarizes representative data-driven models. 

\begin{table*}[!htb]
\caption{Summary of Representative Data-Driven Car Following Models}
\label{tab:data_driven_car_following_models}
\centering 

\begin{tabularx}
{\textwidth}{|>{\columncolor[HTML]{EFEFEF}\centering\arraybackslash}p{0.10\textwidth}|>{\centering\arraybackslash}p{0.2\textwidth}|X|X|}

\hline 
\rowcolor[HTML]{EFEFEF}
\textbf{Model} & \textbf{Authors \& Year} & \multicolumn{1}{c|}{\textbf{Key Strengths}} & \multicolumn{1}{c|}{\textbf{Key Limitations}} \\ \hline
\cellcolor[HTML]{EFEFEF}Machine Learning &

Kehtarnavaz et al.  (1998)  \cite{1998_Kehtarnavaz_Groswold_Miller_Lascoe} 
\newline
Panwai and Dia (2007) \cite{2007_Panwai_Dia}
\newline
Khodayari et al.  (2012)  \cite{2012_Khodayari_Ghaffari_Kazemi_Braunstingl} 
\newline
Zheng et al. (2013) \cite{ 2013_Zheng_Suzuki_Fujita}
\newline
Wei and Liu. (2013) \cite{2013_Wei_Liu} 
\newline
Papathanasopoulou and Antoniou (2015)  \cite{2015_Papathanasopoulou_Antoniou}
\newline
Dabiri and Abbas (2018) \cite{2018_Dabiri_Abbas}
\newline
Yang et al. (2018)  \cite{2018_Yang_Zhu_Liu_Wu_Ran}
\newline
Soldevila et al. (2021) \cite{ 2021_Soldevila}
\newline
Kamjoo et al. (2022)  \cite{2022_Kamjoo_Saedi_Zockaie_Ghamami_Gates_Talebpour}

& 
1.	Interpretability: Traditional ML models (e.g. Linear Regression, Decision Tree and SVM) are interpretable, making it easier to understand how decisions are being made, which is crucial for safety-critical applications. 
\newline
2.	Faster Calibration and Inference: Due to their simpler nature, traditional ML models can be trained faster and often have lower latency during inference. 
&
1.	Feature Engineering: ML models often rely heavily on manual feature engineering.
\newline
2.	Scalability: As the complexity of the scenario increases, the performance of traditional ML models might not scale as well as other models.
\\
\hline 
Deep Learning &
Morton et al. (2016) \cite{2016_Morton_Wheeler_Kochenderfer} 
\newline
Zhou et al. (2017) \cite{2017_Zhou_Qu_Li} 
\newline
Wang et al. (2017) \cite{2017_Wang_Jiang_Li_Lin_Zheng_Wang} 
\newline
Wang et al. (2019) \cite{2019_Wang_Jiang_Li_Lin_Wang} 
\newline
Wu et al. (2019) \cite{2019_Wu_Tan_Chen_Ran}
\newline 
Zhang et al. (2019) \cite{2019_Zhang_Sun_Qi_Sun}
\newline 
Ma and Qu (2020) \cite{2020_Ma_Qu} 
\newline
Zhou et al. (2022) \cite{2022_Zhou_Wan_Zhu} 
\newline
Xu et al. (2023) \cite{2023_Xu_Gao_Qiu_Li} 
\newline
Lu et al. (2023) \cite{2023_Lu_Yi_Liang_Rui_Ran}
\newline
Xu et al. (2024) \cite{2024_Xu_Chen_Zhang_Wang_Liu_Guo}
\newline
Chen et al. (2024) \cite{2024_Chen_Yuan_Zhu_Zheng_Shen_Wang}
&
1.	Data-driven: DL models learn directly from data, allowing them to capture the complexity of real-world driving behavior without the need for explicit modeling assumptions.
\newline
2.	Flexibility: A wide range of DL techniques can be used, enabling the selection of an appropriate model for the specific problem and data. 
\newline
3.	Incorporation of driver characteristics: DL models can learn to capture driver characteristics or latent driving styles, enabling the modeling of heterogeneous human driving behaviors. &
1.	Labeled data requirements: Supervised learning models require a large amount of labeled training data.
\newline
2.	Cascading errors: In sequential prediction tasks, an incorrect estimation at an early stage can propagate through subsequent steps, leading to increasingly inaccurate decisions. 
\\
\hline
Imitation Learning &
Shimosaka et al. (2014) \cite{2014_Shimosaka_Kaneko_Nishi} \newline
Shimosaka et al. (2015) \cite{2015_Shimosaka_Nishi_Sato_Kataoka} 
\newline
Kuefler et al. (2017) \cite{2017_Kuefler_Morton_Wheeler_Kochenderfer} \newline
Gao et al. (2018) \cite{2018_Gao_Shi_Xie_Cheng} \newline
Zhou et al. (2020) \cite{2020_Zhou_Fu_Wang} \newline
Zhao et al. (2022) \cite{2022_Zhao_Wang_Han_Gupta_Tiwari_Wu_Barth}
&
1. Capturing implicit knowledge: Imitation can capture the implicit knowledge and preferences of human drivers, such as safety, comfort, and efficiency, without the need to explicitly specify these objectives.
\newline
2. Robustness: By learning from expert demonstrations, Imitation models can be more robust to environmental uncertainties and disturbances, as the expert's behavior may already account for these factors. &
1. Quality of expert demonstrations: The effectiveness of IL depends on the quality and diversity of the expert demonstrations provided. 
\newline
2. Ambiguity in Learned Policy: It is important to note that if multiple behaviors could result in the same observed behavior, IL might not distinguish the true intent behind the actions.
\\
\hline
Unsupervised Learning &
Koutsopoulos and Farah (2012) \cite{2012_Koutsopoulos_Farah} 
\newline
He, et al. (2015) \cite{2015_He_Zheng_Guan} \newline
Lefevre et al. (2015) \cite{2015_Lefevre_Carvalho_Borrelli}
Guha et al. (2022) \cite{2022_Guha_Lei_Zhu_Nguyen_Zhao}
&
1. Data Efficiency: Unsupervised learning (UL) can utilize large amounts of unlabeled data, which are more abundant and less costly to obtain than labeled data.
\newline
2. Anomaly Detection: UL is particularly good at detecting outliers or anomalies, which could be crucial for identifying unexpected behaviors or conditions.
\newline
3. Clustering Behaviors: UL can cluster similar driving behaviors, which could be used to understand different driving styles and adapt the car-following behavior accordingly.
&
1. Validation Challenges: Without labeled data, it can be challenging to validate the model's predictions or to determine its accuracy in a traditional sense.
\newline
2. Risk of Overfitting: There's a risk that the model will overfit to the noise in the data, detecting spurious patterns that do not generalize well to unseen situations. 
\\
\hline
Reinforcement Learning &
Ioannou and Chien (1993) \cite{1993_Ioannu_Chien}
\newline
Desjardins and Chaib-draa (2011) \cite{2011_Desjardins_Chaib-Draa}
\newline
M. Zhu et al. (2018) \cite{2018_Zhu_Wang_Wang}
\newline
Y. Zhang et al. (2019) \cite{2019_Zhang_Guo_Gao_Qu_Chen}
\newline
M. Li et al. (2020) \cite{2020_Li_Li_Xu_Liu}
\newline
Yen et al. (2020) \cite{2020_Yen_Chou_Shih_Chen_Tsung}
\newline
Masmoudi et al. (2021) \cite{2021_Masmoudi_Friji_Ghazzai_Massoud}
\newline
Yavas et al. (2022) \cite{2022_Yavas_Kumbasar_Ure}
\newline
Wang et al. (2022) \cite{2022_Wang_Huang_Tang_Meng_Hu}
\newline
Tang et al. (2022) \cite{2022_Tang_Chen_Yang_Toyoda_Liu_Hu}
\newline
Liao et al. (2024) \cite{2024_Liao_Yu_Chen_Zhou_Li}

&
1. Adaptability: RL is highly adaptable to a range of conditions and can improve its policy based on the feedback from the environment. 
\newline
2. Optimization: RL agents can learn to optimize multiple objectives simultaneously, balancing trade-offs between safety, efficiency, and comfort. 
\newline
3. Exploration vs. Exploitation: RL allows the algorithm to discover new strategies while still leveraging what has been learned. 
&
1. Training stability: RL algorithms can be sensitive to hyperparameters and prone to instabilities during training, requiring careful tuning and design choices. 
\newline
2. Computational complexity: RL often requires significant computational resources for both training and inference, which can be a limiting factor. 
\newline
3. Safety concerns: RL agents may explore suboptimal or unsafe actions that could lead to accidents or undesirable behaviors during the learning process as a concern in safety-critical applications. \\
\hline

\end{tabularx}
\end{table*}
\subsection{Supervised Learning Models}
\subsubsection{Machine Learning Model}
With the availability of high-resolution vehicle trajectory data and the rampant development of machine learning (ML) algorithms and tools, several high-accuracy machine learning and deep learning algorithms have been developed. Kehtarnavaz et al. \cite{1998_Kehtarnavaz_Groswold_Miller_Lascoe} developed a time-delay neural network (TDNN) autonomous vehicle following module to replace a conventional proportional and derivative/proportional and integral (PD/PI) controller and implemented it on actual vehicles. Panwai and Dia \cite{2007_Panwai_Dia} developed a neural agent car-following model for mapping perceptions to actions. They compared the speed and position of individual vehicles from data to model output trajectory profiles. Khodayari et al. \cite{2012_Khodayari_Ghaffari_Kazemi_Braunstingl} developed an Artificial Neural Network (ANN) to simulate and predict the car-following behavior based on the driver-vehicle unit's reaction delay, and input parameters include relative speed, relative distance, and follower's speed. Zheng et al. \cite{2013_Zheng_Suzuki_Fujita} developed ANN to output the following vehicle’s speed by considering the driver–vehicle reaction delay in relative speed and acceleration, the gap, and the vehicle's speed. Wei and Liu \cite{2013_Wei_Liu} proposed a Support Vector Machine approach to investigate the asymmetric characteristics of car-following behavior and its impact on traffic flow evolution. The Support Vector Machine model takes space headway, velocity, and relative speed as the inputs and the output is the follower's speed. Another study \cite{2015_Papathanasopoulou_Antoniou} proposed a data-driven locally weighted regression approach for optimizing car-following model estimation by applying an existing technique and calibrated using instrumented vehicles in Naples, Italy.  The gradient boosting of regression tree (GBRT) algorithm \cite{2018_Dabiri_Abbas} was applied to NGSIM data to develop a car-following model that outperforms traditional GHR models in capturing the motion characteristics of successive vehicles, using cross-validation and sensitivity analysis for parameter tuning.
Yang et al. \cite{2018_Yang_Zhu_Liu_Wu_Ran} developed a hybrid method by integrating theoretical-based and machine-learning models to obtain better accuracy while maintaining the physical meaning. Kamjoo et al. A Gaussian Process Regression Approach \cite{2021_Soldevila} to model individual longitudinal driving behaviors by integrating parametric and non-parametric methods, enhancing predictions with large data sets and adapting to new variables without changing the model structure.  \cite{2022_Kamjoo_Saedi_Zockaie_Ghamami_Gates_Talebpour} implemented Support Vector Machine and Long-Short-Term-Memory (LSTM) models to investigate the winter operation's impact on car-following behavior, showing that the presence of snowplows leads to significantly different car-following behaviors.
\subsubsection{Deep Learning Model}
Driving behaviors can be treated as sequential data since the driver's following action is conditioned on their previous action. Deep Learning (DL) models, particularly Recurrent Neural Networks (RNN), are similar to humans in memory-based decision-making. Morton et al. \cite{2016_Morton_Wheeler_Kochenderfer} evaluated LSTM networks for simulating driver acceleration behavior on highways, which can effectively replicate real driver behavior in traffic models, matching or surpassing baseline methods in accuracy and realism. Zhou et al. \cite{2017_Zhou_Qu_Li} developed an RNN with inputs of the gap, relative speed, and follower's speed to predict acceleration, speed, and gap that can be used for car-following control. Wang et al. (2017) \cite{2017_Wang_Jiang_Li_Lin_Zheng_Wang} developed deep learning-based car-following with Gated Recurrent Unit (GRU) that embeds model outputs' prediction and memory effects.  Wang et al. (2019) \cite{2019_Wang_Jiang_Li_Lin_Wang} applied Gated Recurrent Unit (GRU) with different timescale historical information as inputs to test memory effects on car following behavior. Wu et al. (2019) \cite{2019_Wu_Tan_Chen_Ran} built a deep learning-based model to mimic human drivers' memory, attention, and prediction (MAP) mechanisms. Zhang et al. \cite{2019_Zhang_Sun_Qi_Sun} built an LSTM model with a hybrid retraining constrained (HRC) training method to model car-following and lane-changing behaviors. Given that CF behaviors (acceleration, deceleration, and cruising) are made for continuous time steps rather than step-by-step with memory effects and delayed reaction time, Ma and Qu \cite{2020_Ma_Qu} developed a seq2seq LSTM model for predicting multistep car-following behaviors. Their model inputs include gap distance, relative speed difference, and the speed of the subject vehicle, and the model output is the multiple-step accelerations/decelerations. Zhou et al. \cite{2022_Zhou_Wan_Zhu} applied a transfer learning-based LSTM car-following model for adaptive cruise control to address the ACC data scarcity problem by transferring useful features from human-driven data. Missing data in vehicle state information often leads to gaps in the data required for making accurate control decisions. A longitudinal CF model leverages Transformer-Generative Adversarial Networks (TransGAN) \cite{2023_Xu_Gao_Qiu_Li} for this purpose, which is particularly significant in scenarios where data acquisition devices malfunction. 
An improved sequence-to-sequence model for connected automated vehicles (CAVs) \cite{2023_Lu_Yi_Liang_Rui_Ran} enhances the accuracy of simulating speeds and positions by incorporating kinematics information from multiple preceding vehicles and using a novel deep learning framework. Another sequence-to-sequence model \cite{2024_Xu_Chen_Zhang_Wang_Liu_Guo} incorporates an attention mechanism to learn the probability distribution of driver reaction delay. Aggressiveness Informed Car-Following (AICF) \cite{2024_Chen_Yuan_Zhu_Zheng_Shen_Wang} was proposed approach, which integrates driving style as a dynamic input in data-driven car-following models to address the limitations of using predefined patterns and categories.

\subsubsection{Imitation Learning Models}
Imitation learning (IL) a subset of machine learning focused on learning from demonstration, which has several advantages: 1. It can meet the preferences of driving styles; 2. Other drivers would easily understand the automated car-following maneuver. Behavior Cloning (BC) models, which are the simplest forms of Imitation Learning, try to learn the maneuver of the human driver given the expert’s demonstrations that are divided into state-action pairs. The BC model can discover complex car-following strategies for nonlinear and high-dimensional systems. Inverse reinforcement learning (IRL) can be seen as a form of imitation learning where the model is not just learning to mimic the actions of the expert, but is actually trying to understand the motivations (i.e., the reward function) behind those actions. This can lead to more robust and versatile models, as they are not just copying behavior, but are learning the principles that guide that behavior.  With naturalistic driving behavior data, the IRL would approximate that human's reward function for the driving task. Instead of learning a policy from a reward function, Inverse Reinforcement Learning (IRL) is trying to learn a reward function from the demonstration of a task \cite{2000_Ng_Russell}. Shimosaka et al. \cite{2014_Shimosaka_Kaneko_Nishi} proposed an IRL framework to model risk anticipation and defensive driving, which has a better long-term predict \cite{2015_Shimosaka_Nishi_Sato_Kataoka} developed multiple reward functions with clustering of environments to better handle environmental diversity. The framework then simultaneously learns the parameters specific to each cluster. This method significantly improves the model's adaptability and accuracy in diverse settings. Kuefler et al. \cite{2017_Kuefler_Morton_Wheeler_Kochenderfer} proposed Generative Adversarial Imitation Learning (GAIL) framework to imitate human car-following behavior with three non-linearity in mapping from states to actions, high-dimensional state representation, and stochasticity.  Gao et al. \cite{2018_Gao_Shi_Xie_Cheng} proposed a car-following algorithm with a reward function based on IRL under different conditions. The experimental verification is conducted on a dynamic driving simulation test bench. Zhou \cite{2020_Zhou_Fu_Wang} developed a Maximum Entropy Deep IRL approach using a neural network to estimate the rewards. Model input features include time headway, relative speed, and maximum speed, and model output is speed control. Zhao et al. \cite{2022_Zhao_Wang_Han_Gupta_Tiwari_Wu_Barth} developed a Personalized-ACC framework using IRL, which classified the driver and weather conditions for the pre-trained off-line IRL model and compared it with the IDM model regarding safety and comfort measured by takeover percentage. 

\subsection{Unsupervised Learning Models} Unlike traditional approaches that rely on supervised learning with labeled datasets, unsupervised learning based car following models are designed to autonomously discover patterns and relationships within unlabeled traffic data. This paradigm shift facilitates a more organic understanding of traffic dynamics. As vehicular technologies and traffic systems evolve towards greater autonomy and complexity, unsupervised learning provides a crucial framework for developing more adaptive, efficient, and safer car following algorithms. A latent class model \cite{2012_Koutsopoulos_Farah} is developed to estimate drivers' decision, based on the assumption that these decisions are probabilistic and the result of a number of explanatory variables. The latent class model is formulated as a discrete choice problem with systematic utility functions. A simplified k-nearest mean model \cite{2015_He_Zheng_Guan} was developed that only takes the position as input and reproduces traffic dynamics without premises and calibration. This approach selects the most similar historical records and outputs the average of the k nearest neighbors. A learning-based architecture for autonomous driving which combines a driver model and a Model Predictive Controller (MPC) \cite{2015_Lefevre_Carvalho_Borrelli}. The driver model is built on Hidden Markov models (HMM) representation of human control strategies to imitate different driving styles. The MPC guarantees safety by enforcing a set of constraints on the vehicle state. A robust unsupervised learning method \cite{2022_Guha_Lei_Zhu_Nguyen_Zhao} is proposed to model temporal dynamic interaction between two agents comprises a pair of well-aligned trajectories, which provides more robust and data-driven learning approaches for highly diverse and dynamic interactions. The unsupervised learning framework doesn't rely on strong modeling assumptions that are often used to represent latent variables that govern the actions among agents and the driving environment. 

\subsection{Reinforcement Learning}
In the development of advanced car-following models, Reinforcement Learning (RL) has emerged as an important approach due to its capacity for decision-making in complex and dynamic environments. RL can be broadly classified based on the knowledge of the environment: model-based RL, where the model of the environment is known, and model-free RL, where it is not. Furthermore, RL methods can also be categorized based on their use of the optimality condition to derive an optimal policy. Indirect RL methods leverage solutions from the Bellman or Hamilton-Jacobi-Bellman (HJB) equations, founded on Bellman’s principle of optimality, to ascertain at least one optimal policy. In contrast, direct RL methods avoid the optimality condition, opting instead to search the entire policy space to optimize the primal problem directly. By utilizing these RL strategies, car-following models aim to replicate and optimize human-like driving behaviors, focusing on safety, efficiency, and comfort while navigating the uncertainties of real-world traffic conditions. The RL agent chooses an action $a$ from action space $A$ based on the observed state $s$ to get a reward $R(s,a)$. During the learning process, the agent uses an exploration-exploitation trade-off. It explores the environment by taking random actions to discover new states and learn more about their rewards. The agent gradually learns the optimal action policy for each state by repeatedly updating the Q-values based on observed rewards. 
\begin{equation}
\label{deqn_ex1a}
\begin{split}
\text{New } Q\left(s,a\right) & \gets Q\left(s,a\right) \\
& \quad + \alpha \left[ R\left(s,a\right) + \gamma \max Q^\prime(s^\prime,a^\prime) - Q(s,a) \right]
\end{split}
\end{equation}
where $\alpha$ is the learning rate, and $\gamma$ is the discount rate. $Q^\prime(s^\prime,a^\prime)$ is next Q-value for all possible $s^\prime$ and $a^\prime$. $R\left(s,a\right)$ is the reward function. 

Ioannou and Chien \cite{1993_Ioannu_Chien} applied RL to develop an autonomous intelligent adaptive cruise control system. Desjardins and Chaib-draa \cite{2011_Desjardins_Chaib-Draa} introduced a policy gradient algorithm estimation and used BPNN for longitudinal vehicle control. Their RL-based car-following control was implemented in the simulator to realize the efficient behavior of CACC. Zhu et al. \cite{2018_Zhu_Wang_Wang} proposed a framework for developing human-like car-following based on deep reinforcement learning using speed deviations as a reward function. Zhang et al. \cite{2019_Zhang_Guo_Gao_Qu_Chen} developed a deterministic promotion RL algorithm (DPRL) based on an actor-critic frame and the policy gradient method for longitudinal velocity control, implemented in CarSim simulated environment and real-world experiment. Li et al. \cite{2020_Li_Li_Xu_Liu} proposed a deep deterministic policy gradient (DDPG)-based driving strategy taking information from in-vehicle sensors for system performance optimization. Yen et al. \cite{2020_Yen_Chou_Shih_Chen_Tsung} developed a proactive car-following model using deep-RL considering road efficiency, safety, and comfort, which shows better performances in terms of time heady, time to collision (TTC), and jerk (rate of change in acceleration). Yavas et al. \cite{2022_Yavas_Kumbasar_Ure} developed a Model-Based Reinforcement Learning (MBRL) for ACC control with a hybrid policy that combines the Intelligent Driver Model following policy with the Deep Reinforcement Learning policy. Wang et al. \cite{2022_Wang_Huang_Tang_Meng_Hu} devised an RL model for longitudinal velocity control in a three-vehicle mode based on the safety level, consisting of the collision avoidance priority of the leading and following vehicles and the expected acceleration/deceleration decision. Most CF models take relative speed, acceleration, and gap as inputs, while some studies \cite{2021_Masmoudi_Friji_Ghazzai_Massoud, 2022_Tang_Chen_Yang_Toyoda_Liu_Hu} use video object detection to learn the driving strategy. Such models demonstrate promising results, which can efficiently be deployed on embedded devices for advanced driver assistance systems. This study \cite{2024_Liao_Yu_Chen_Zhou_Li} develops a personalized car-following model using a memory-based deep reinforcement learning approach, combining Twin Delayed Deep Deterministic Policy Gradients (TD3) with Long Short-Term Memory (LSTM) networks to optimize car-following behavior based on diverse driving styles for improved safety, efficiency, and comfort, demonstrating superior performance and convergence compared to other models.


\section{Knowledge-driven Models}
Knowledge-driven methods leverage extensive pre-trained Large GenAI models and foundational knowledge to generate new content and solutions, simulating commonsense-based understanding and reasoning that reflect deep concepts. Large GenAI models can address the heterogeneity in traffic components, varying road environments, and the stochastic nature of driving with an agent-based approach. CF behavior is considered as a subtask of Autonomous Driving (AD), which suggests that large GenAI models developed in AD can also be applicable in CF applications. The kinematic CF models (e.g., IDM models) are useful as baseline for validating new knowledge-based driving agents. In CF scenarios, these knowledge-driven methods reason about vehicle behaviors, enhancing decision-making in complex driving scenarios. Additionally, they provide context-aware systems that understand and react to real-time driving conditions, leading to more realistic simulations, optimal control, and enhanced human-machine interaction. 

\subsection{Large Language Models (LLMs)}
LLMs can generate detailed driving instructions and responses based on signal inputs and human-like understanding. These models can create complex driving policies by predicting and generating the next action based on previous driving scenarios and contextual data, which enhances the decision-making processes in autonomous CF, making vehicles more adaptive and responsive to dynamic driving environments. Examples of LLM-based autonomous driving models include DriveLLM \cite{2023_Cui_Huang_Zhong}, DiLu \cite{2023_Wen_DiLu}, Driving with LLMs \cite{2023_Chen_Sinavski}, and Drive-as-you-speak \cite{2024_Cui_Yang}.

\subsection{Vision-Language Models (VLMs)} 
VLM models integrate visual inputs, such as images and videos, with textual descriptions to enhance understanding and generate new insights or content. These models excel in scenarios that require multi-modal perception and comprehension, enabling accurate interpretations of complex driving situations. Representative models include vision-language-action driving model (VLAM) (LINGO-1 \cite{2023_Wayve} \& LINGO-2 \cite{2024_Wayve}) and DriveVLM \cite{2024_Tian}.

\subsection{Multimodal Large Language Models (MLLMs)} 
MLLMs extend the capabilities of traditional LLMs and VLMs by integrating multiple types of data inputs. These models can process and understand a combination of various sensor input to provide a holistic view of the driving environment. A recent survey paper \cite{2023_Cui_Ma} on Multimodal Large Language Models for AD sheds light on the latest advancements and research directions in the field. Representative models include DriveMLM \cite{2023_Wang_Xie_Hu} and LMDrive \cite{2023_Shao}. 

\section{Discussion}
\begin{figure*}[!htbp]
  \centering 
  \includegraphics[width=\textwidth]{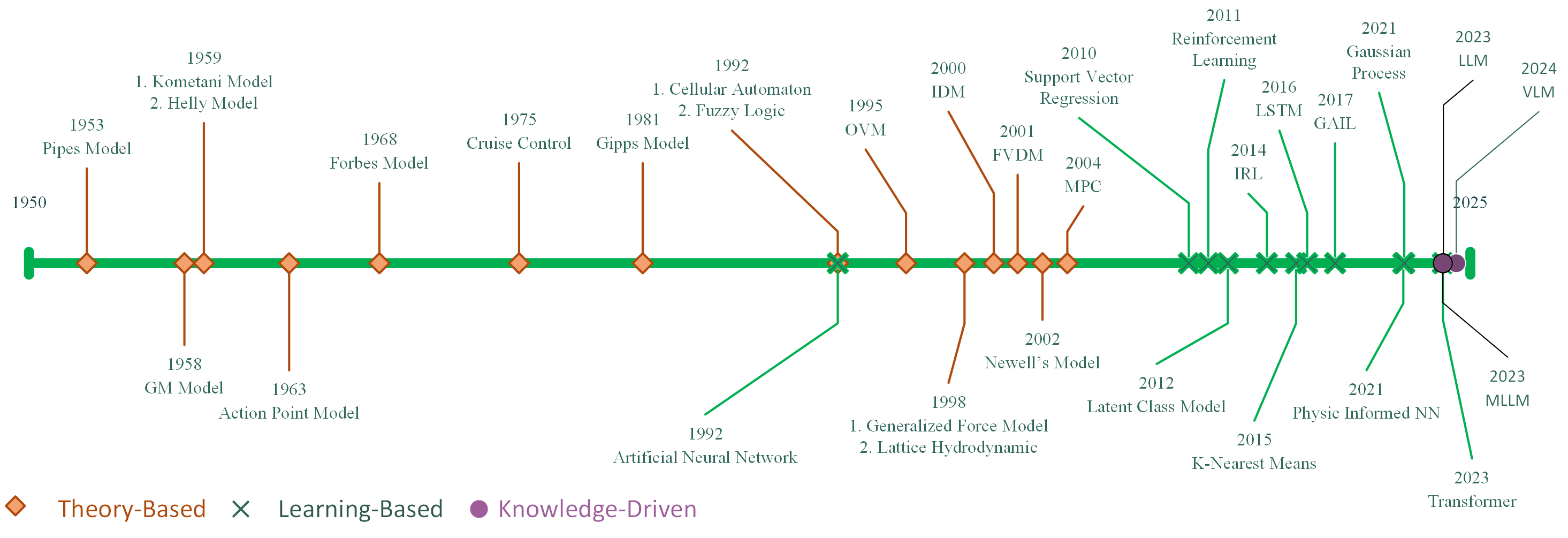}
  \caption{The Historical Timeline of Reviewed Master Models and Key Variants.}
  \label{fig_timeline}
\end{figure*}

\begin{figure*}[!htbp]
  \centering 
  \includegraphics[width=1.0\textwidth]{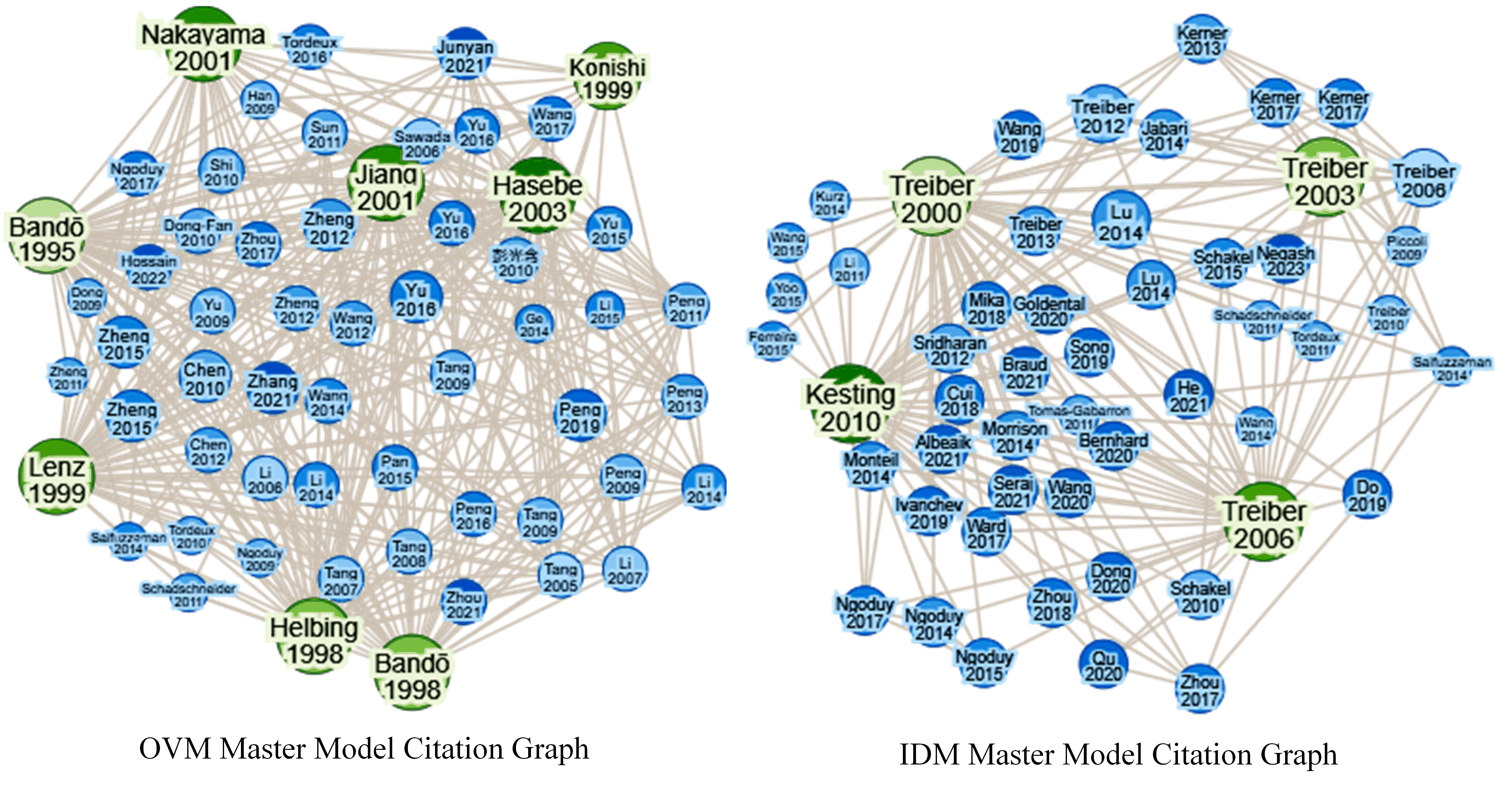}
  \caption{Navigating the Literature with Reviewed Master Models Using OVM and IDM as Examples}
  \label{fig_citation_relation}
\end{figure*}

In this review paper, the focus on master models (Figure \ref{fig_timeline}) offers a structured approach by providing a clear framework for categorization and understanding. The Figure \ref{fig_citation_relation} illustrates the efficacy of these master models in delineating the progression and branching out of various derivative algorithms within the field. By doing so, it highlights the master models' role as pivotal reference points from which the landscape of car-following behavior can be surveyed. Automatic literature tools \cite{2023_research_rabbit} adeptly navigate the extensive body of literature to surface the most relevant studies, demonstrating the significant advantage of leveraging structured categorizations to manage and interpret the wealth of research in this evolving area of traffic dynamics and control systems. This paper's focused review through the lens of multi-disciplines thus not only consolidates existing knowledge but also guides future investigations.

\subsection{Theory-Based vs. Learning-Based vs. Knowledge-Driven}
Decades of theory-based driving behavior research are now merging with learning-based and knowledge-driven opportunities. Although the data-driven models will undoubtedly become critically important, the traditional driver behavior expertise remains relevant. It is auspicious to integrate rule-based models with the fast-evolving AI models, which can reason, learn, and fill in missing information to achieve better interpretability, generalizability, and reliability.

\begin{table*}[tbhp]
\centering
\caption{Comparative Overview of Theory-Based Models vs Learning-Based Models vs Knowledge-Driven Models}
\label{tab:theory_compare_data}
\fbox{ 
\begin{minipage}{\textwidth} 
\begin{tabularx}{\textwidth}{X X X}
\toprule
\textbf{Theory-Based Models} & \textbf{Learning-Based Models} & \textbf{Knowledge-Driven Models} \\
\midrule
\textbf{Strengths} & \textbf{Strengths} & \textbf{Strengths} \\
\midrule
\begin{itemize}
    \item Solid Theoretical Foundation: Built on well-established principles and theories, offering predictable behavior within their domain of validity.
    \item Interpretability: Due to their reliance on explicit formulas and parameters, these models provide clear insights into the cause-and-effect relationships within the modeled phenomena.
    \item Generalizability: Capable of generalizing across different scenarios, ensuring that the model adheres to expected physical laws and behavioral norms, which helps in maintaining performance even under unusual conditions not covered by the data.
\end{itemize}
&
\begin{itemize}
    \item Accuracy and Prediction: Learning-Driven models are adept at capturing non-linear patterns from large datasets. This can lead to highly accurate predictions based on real-world data. 
    \item Adaptability to Complex Environments: Ability to achieve human-level or superior performances under diverse driving conditions.
    \item Handling of Heterogeneous Data: Efficient processing and learning from varied data inputs, adapting to different traffic and infrastructure conditions.
\end{itemize}
&
\begin{itemize}
    \item Physical-Social Intelligence: They understand both physical laws and drivers' expectations and emotional needs, enabling them to identify and comprehend various patterns, rules, and relationships in driving behavior. 
    
    \item Handling Edge Cases: Knowledge-Driven Large GenAI models address edge cases and unusual scenarios by leveraging commonsense reasoning and proactive decision-making.
    
    \item Context-Aware Systems: They enable context-aware systems that understand and react to real-time driving conditions, enhancing decision-making, human-machine interaction, and the overall safety and efficiency of intelligent vehicles.
\end{itemize}
\\
\midrule
\textbf{Limitations} & \textbf{Limitations} & \textbf{Limitations} \\
\midrule
\begin{itemize}
    \item Complexity of Traffic Dynamics: Struggles to account for all heterogeneous traffic compositions, infrastructure conditions, and varying data inputs.
    \item Parametric Restrictions: Limited by the need to pre-define parameters, which may not capture the full spectrum of driver behaviors or environmental conditions.
\end{itemize}
&
\begin{itemize}
    \item Overfitting and Scalability: Due to environmental diversity and sensor variability, learning-based techniques perform well on the data on which they were trained but not always adaptable on new, unseen datasets or in more complex environments.
    \item Learning Undesirable Behaviors: Potential to learn from bad behaviors due to drivers' unpredictability or poor data quality.
\end{itemize}
&
\begin{itemize}
    \item Alignment Challenges: Knowledge-Driven pre-trained foundational models are end-to-end systems. Misalignment occurs when objectives and behavior do not align with the intended use case, leading to undesired consequences or vulnerabilities.
    \item Safety and Ethical Concerns: Addressing potential biases, prioritize safety, and ensuring transparency in decision-making, are ongoing challenges that need careful consideration.
\end{itemize}
\\
\bottomrule
\end{tabularx}
\end{minipage} 
} 
\end{table*}

Several hybrid CF models were developed by integrating rule-based and AI-based models, which leverages the strengths of different approaches, enhancing accuracy, robustness, and safety in systems like autonomous vehicle control, where both predictable behavior and adaptability to real-world complexity are crucial. Li et al. \cite{2019_Li_Lu_Ren_Zhao} utilized the adaptive Kalman filter algorithm to integrate the LSTM data-driven model with the IDM theoretical-driven model to build the car-following model. Mo et al. \cite{2021_Mo_Shi_Di} designed Physic-Informed Deep Learning architectures encoded with IDM and OVM to predict accelerations in four traffic regimes: acceleration, deceleration, cruising, and emergency braking. This paper \cite{2023_Shiomi_Li_Knoop} focused on analyzing the stochastics and heterogeneity in traffic dynamics using a data-driven car-following model. It examined the shape of the distribution of a stochastic term in the car-following model, predicting accelerations accurately and identifying relationships between predicted distributions' modality, standard deviation, and traffic states. This research \cite{2023_Pan_Zhang_Tian_Cui_Wang} introduces the IDM-Bi-LSTM combination car-following model, which integrates the theory-driven Intelligent Driver Model (IDM) with a data-driven model optimized by Particle Swarm Optimization (PSO) to enhance trajectory prediction accuracy and maintain controllability and security. Numerical simulations demonstrate that this hybrid model significantly reduces prediction errors and outperforms both individual IDM and PSO-Bi-LSTM models in simulating heterogeneous and asymmetric driving behaviors. 

The comparison emphasizes (Table \ref{tab:theory_compare_data}) the contrast between Rule-based and AI-based approaches in modeling car-following behavior. While theory-based models offer a clear, predictable framework grounded in established principles, their limitations in handling the complexity and variability of real-world driving scenarios highlight the potential benefits of incorporating learning and commonsense knowledge. Conversely, AI-driven models' ability to adapt to and learn from diverse environments is tempered by challenges related to interpretability, the quality of learned behavior, and the handling of rare but critical situations. Balancing these strengths and limitations often results in more reliable, accurate, and practical systems for real-world applications.

\subsection{Applications}

Analyzing car-following behavior and developing intelligent control algorithms have significant consequences for road safety and energy efficiency. The profound impacts of the different driving algorithms (e.g., strategic models, model-predictive control, RL controller, and LLM-agent) and various connected environments have set the stage for a new era of traffic flow dynamics. Many applications based on car-following models have been undertaken, summarized with the following topics.


\begin{itemize}
    \item \textbf{Topic 1:} Traffic Flow Dynamics. Simulation analysis through car following models offer insights into emerging traffic flow characteristics under conventional or connected environment. CF models serve as the basis for exploring how connected and automated mobility (CAM) could enhance traffic dynamics and address both the opportunities and challenges presented by mixed traffic scenarios \cite{2020_Chen_Srivastava_Ahn_Li,2021_Kontar_Li_Srivastava_Zhou_Chen_Ahn,2018_Ye_Yamamoto,2015_Li_Zhang_Zheng_He_Peeta_Zheng_Li,2018_Zhu_Zhang,2016_Talebpour_Mahmassani,2018_Rahman_Abdel-Aty,2023_Huang_Ye_Sun_Tian,2018_Sun_Zheng_Sun,2020_An_Xu_Qian_Chen_Luo_Li,2018_Sun_Chen_Zhao_Liu_Zheng, 2022_Ma_Wu_Hu_Chen_Luo, 2023_Kamjoo_Saedi_Zockaie_Ghamami_Gates_Talebpour}. 
    \item \textbf{Topic 2:} Road Safety and Energy Efficiency. Car-following models play a crucial role in developing eco-driving strategies and improving safety levels on the road. By simulating the behavior of vehicles in various traffic conditions, these models can help identify optimal driving patterns that minimize energy consumption and emissions while also reducing the risk of accidents. As research in this area continues to evolve, more sophisticated systems will be evaluated and developed with CF models to enhance both environmental sustainability and road safety.\cite{2023_Garg_Bouroche, 2022_Krishnamurthy_Huang, 2021_Fleming_Yan_Lot,2015_Li_Xu_Huang_Cheng_Peng,2017_Shengbo_Eben_Li,2018_Xin_Fu_Yuan_Liu_Yu} 
    \item \textbf{Topic 3:} V2X and Wireless Communication. Car-following models can significantly enhance wireless communication applications within the Internet of Vehicles (IoV) to improve situational awareness. CF models facilitate the development of algorithms that optimize data transmission and reception among connected vehicles, improving the efficiency of vehicular ad-hoc networks (VANETs). This fosters the deployment of advanced IoV services, and the delivery of various Internet-based applications to drivers and vulnerable road users (VRU), thereby leveraging the full potential of cloud computing and IoT technologies in the automotive domain  \cite{2012_Rawashdeh_Mahmud, 2020_Sun_Huang_Zhang,2020_Li_Hui_Zhao_Liu_Ngoduy,2016_Jia_Ngoduy,2021_Li_Han_Ma,2019_Navas_Milanés,2017_Meng_Li_Wang_Li_Li,2016_Jia_Ngoduy_2}. 
    \item \textbf{Topic 4:} Innovative Traffic Control. Researchers have focused on traffic control and novel operation strategies. CF models are employed to optimize the control algorithms, such as collaborative driving strategy, traffic signal timing, shock-wave damping, or variable speed limits \cite{2015_Taniguchi_Nishi_Ezaki_Nishinari,2020_Nishi,2008_Kesting_Treiber_Schönhof_Helbing,2018_Stern_Cui_DelleMonache_Bhadani_Bunting_Churchill_Hamilton_Pohlmann_Wu_Piccoli_Seibold,2017_Cui_Seibold_Stern_Work,2021_Chen_Wang_Gong_Zhou_Ran,2023_Wang_Jin,2018_Vinitsky_Kreidieh_LeFlem_Kheterpal_Jang_Wu_Wu_Liaw_Liang_Bayen, 2023_Fu_Keridieh_Wang_Lee_Delle_Bayen, 2023_Amaury}. 
    \item \textbf{Topic 5:} Cyber-Physical Human Systems. The seamless integration of human elements, computer-based algorithms, and physical components is essential to transform transportation systems into Cyber-Physical Human Systems (CPHS). By leveraging car following models, researchers are able to develop analytical frameworks that not only enhance the performance of these vehicles but also improve the overall security of the transportation system. This integration and testing pave the way for the eventual realization of fully autonomous driving capabilities \cite{2017_Hu_Wang_Tang, 2020_Jin_Sun_Zhao, 2020_Wang_Wu_He,2019_Wang_Yu_Wu_Wang_He,2021_Nice_Elmadani, 2021_Wang_Han_Tiwari, 2022_Sean_McQuade, 2024_Li_Jin, 2024_Raveendran_Mathew_Velaga}. 
    \item \textbf{Topic 6:} Human Factors Research. Car following models are effective tools to incorporate psychological and behavioral aspects to assess human-machine interaction, driver behavior, and the potential consequences of automated systems. By understanding human elements and studying human-machine interfaces, researchers aim to improve the acceptance, trust, and performance of intelligent vehicles \cite{2015_Saifuzzaman_Zheng_Haque_Washington,2020_Chen_Sun_Ma_Sun_Zheng,2017_Ro_Roop_Malik_Ranjitkar,2018_VanLint_Calvert,2022_Harth_Amjad_Kates_Bogenberger,2018_Huang_Sun_Sun,2019_Hoel_Driggs-Campbell_Wolff_Laine_Kochenderfer}.
    \item \textbf{Topic 7:} Trajectory Reconstruction. Sensor-based trajectory acquisition often encounters issues with noise and missing information. This can be due to various factors such as sensor limitations (fixed location or penetration rates), environmental conditions, and the inherent complexity of traffic environments. This is where car following models become crucial to estimate the missing information in vehicle trajectories with the partial data collected from these sensors. The complete and continuous vehicle trajectory is vital for various applications.  \cite{2017_Sazara_Nezafat_Cetin, 2015_Montanino_Punzo, 2022_Chen_Yin_Qin_Tang_Wang_Sun, 2022_Chen_Yin_Tang_Tian_Sun, 2020_Wang_Wei_Chen, 2020_Wei_Wang_Chen, 2020_Lin_Wang_Zhou_Ding_Wang_Tan, 2022_Arman_Tampère, 2023_Makridis_Kouvelas, 2018_Shan_Hao_Chen_Boriboonsomsin_Wu_Barth, 2015_Feng_Sun_Chen, 2023_Zhao_Yang_Zhang}.  
\end{itemize}

\subsection{Dataset}

Detailed high-quality vehicle trajectory data are of great significance for the development of driving behavior models and vehicle control algorithms. With realistic microscopic longitudinal and lateral maneuvers, the valuable vehicle trajectory data lay the foundation for many scientific discoveries. The collection of realistic vehicle trajectory data is identified as important as modeling and understanding the interactions of travelers. There are two types of vehicle driving datasets that are often used as benchmarks. 1. Lagrangian-type data collected from mobile devices on floating vehicles. 2. Eulerian-type traffic-flow data that monitors the entire traffic flow from stationary devices. An integrated Benchmark \cite{2019_Chen_Zhu_Chen_Wang_Lu_Zhong_Han_Wang_Wang} for Car-Following Behavior Modeling is extracted from five public driving datasets under the same format and criteria, which consists of more than 80K car following events. A summary table of datasets suitable for car following model research is presented in Table \ref{tab:car_following_models_dataset}. It is important to note that while this table showcases some of the most pivotal datasets, it is by no means exhaustive. There are undoubtedly other datasets available that could provide valuable contributions to the field. The datasets highlighted in our summary are primarily focused on open street environments capturing captures the complexity and unpredictability of naturalistic driving scenarios as opposed to closed testing environments. 

\begin{table*}[!htbp]
\caption{Summary of Trajectory Dataset Collected from both Mobile and Fixed Sensors}
\label{tab:car_following_models_dataset}
\centering 
\begin{tabularx}{\textwidth}{|
    >{\columncolor[HTML]{EFEFEF}\centering\arraybackslash}p{0.09\textwidth}|
    X|
    >{\centering\arraybackslash}p{0.40\textwidth}|
    >{\centering\arraybackslash}p{0.38\textwidth}|}
    
\hline 
\rowcolor[HTML]{EFEFEF}
\textbf{Dataset} & \textbf{Year}  & \textbf{Highlight} & \textbf{Sample Size} \\ \hline
Next Generation Simulation (NGSIM) \cite{2024_Zhang_Jin_Piccoli_Sartipi} & 2006 & The NGSIM datasets provide an unprecedented depth of trajectory data, which were captured at four locations: The US-101 and I-80 sites offer insights into freeway conditions, while Peachtree Street and Lankershim Boulevard provide data on urban street environments. & Each dataset is segmented into three 15 minute periods, covering approximately 500 meters (1,600 feet) in length, with over 11.8 million rows of trajectory data.  \\ \hline
Zen Traffic Data \cite{2020_Dahiya} & 2018 &  Zen Dataset is characterized by its high traffic volume, two lanes in each direction, and a complex road layout featuring S-shaped curves and elevation changes.  & 3600 trajectories on 2km highway segment for 1 hour.  \\ \hline
HighD \cite{2018_Krajewski} & 2018 & The HighD dataset uses drone on German highways over various sections of highways. & 110,500 vehicles, 28,000 vehicle miles traveled, and 147 hours \\ \hline
ACFR five roundabouts dataset \cite{2019_Zyner} & 2019 & This dataset is aimed at studying naturalistic driving behaviors at unsignalized intersections. The dataset offers insights into a wide range of driving behaviors, such as negligence, speeding, and assertiveness. & A dataset containing over 23,000 vehicle recordings at five different roundabouts, captured by a LiDAR-based detection and tracking system. \\ \hline
INTERA- -CTION \cite{2019_Zhan_Sun_Wang_Shi} & 2019  & The INTERACTION dataset stands out for complex and critical driving situations, often within unstructured settings lacking clear right-of-way rules, where driver behavior is significantly influenced by the actions of others. & Roundabout Scenarios (10,479 vehicle trajectories). Unsignalized intersection scenarios (14,867 trajectories). The merging and lane change scenarios (10,933 trajectories). The signalized intersection (3,775 trajectories). \\ \hline
inD \cite{2019_Bock_Krajewski} & 2020 & The inD (intersection Drone) dataset is a collection of naturalistic road user trajectories at urban intersections. & More than 11,500 road users, including vehicles, bicyclists and pedestrians at intersections with 10 hours of data from four intersections \\ \hline
rounD \cite{2020_Krajewski_Moers_Bock_Vater} & 2020 & The rounD dataset is a collection of road user trajectories at roundabouts in Germany, captured using a drone camera. & A total of 6 hours of recordings with more than 13,746 road users including cars, vans, trucks, buses, pedestrians, bicycles, and motorcycles. \\ \hline
I-24 Motion \cite{2023_Gloudemans} & 2020  & I-24 Motion stands out for its large-scale data acquisition, capturing detailed vehicle trajectories across multiple lanes and over an extended period. & With 276 cameras that cover 4.2 miles of I-24, the data collected is intended to support a wide range of applications.  \\ \hline
pNEUMA \cite{2020_Barmpounakis_Geroliminis} & 2020 & Coordinated Drone video-based vehicle trajectory collected in downtown area in Athens, Greece during morning peak hour. & The pNEUMA dataset encompasses close to half a million vehicle trajectories over a congested 1.3 km² area, featuring more than 100 km-lanes or road network.  \\ \hline
Ubiquitous Traffic Eyes \cite{2021_UTE} & 2021 & The database encompasses a wide array of road conditions and traffic conditions, with manual corrections ensured 100\% vehicle detection and tracking accuracy.  & The data segments vary in duration, ranging from 4 to 20 minutes. This dataset is characterized by its high precision, with time accuracy up to 0.1 seconds and position accuracy up to 0.01 meters. \\ \hline
HIGH-SIM \cite{2021_Shi_Zhao_Yao_Li} & 2021 & High-Granularity Highway Simulation (HIGH-SIM) consists of 2-hour 30fps vehicle trajectory dataset captured by three 8K cameras on the helicopter. & Aerial videos at multiple successive locations of Interstate-75 over 2,438m highway segment, containing data of three lanes and one off-ramp from 4:15 p.m. to 6:15 p.m. \\ \hline
AUTOMATUM DATA \cite{2021_Spannaus} & 2021 & AUTOMATUM Drone-based dataset includes a validation mechanism where trajectories of objects are compared against test vehicles. & Generated from 30 hours video with 18,724 vehicle miles traveled, where the relative speed error is less than 0.2 percent \\ \hline
MAGIC \cite{2021_Ma_Wang} & 2022 & Coordinated Drone video-based vehicle trajectory collected in Shanghai, China during morning peak hour & 3 hours of aerial video at six points on 4,000 m of urban highway (2,000 m in each direction) with four main lanes, three off-ramps, and three on-ramps. \\ \hline
CitySim \cite{2023_Zheng_Aty} & 2022 & CitySim is Drone video-based vehicle trajectory data for safety-oriented research, providing seven bounding box points for fine-grained analysis, with the capability to facilitate digital-twin-related research. & The CitySim dataset provides more than 1,200 minutes of drone video data from multiple countries in diverse weather conditions and locations.  \\ \hline
Safety Pilot Model Deployment (SPMD) \cite{2017_Huang_Zhao_Peng} & 2012 & The SPMD program contains Basic safety messages (BSM), vehicle trajectory, driver-vehicle interaction, and weather data are collected to evaluate V2V communication. & The SPMD data sets contain sanitized mobility data elements collected from about 3,000 vehicles, equipped with connected vehicle technologies, traversing Ann Arbor, Michigan.  \\ \hline
Naturalistic Driving Study (NDS) \cite{2015_Blatt_Pierowicz}  & 2013 & This dataset was collected through an extensive observational study with advanced data acquisition systems, including CAN-Bus, GPS, outward and inward facing cameras, and forward radar to capture driver behavior, vehicle movement, and environmental conditions. & More than 5 million trip files from nearly 3500 drivers located in six geographic regions across the United States.  \\ \hline
nuScenes \cite{2020_Caesar_Bankiti_Lang_Vora} & 2020 & The nuScenes dataset incorporates 6 cameras, 5 radars, and 1 LiDAR sensor. The nuScenes dataset is notable for its holistic view of detailed maps, weather conditions, and day/night cycles. & With almost 1.4 million 3D bounding boxes across 23 object classes and over 8 million annotated instances, the nuScenes dataset spans over 100 hours of driving data, capturing various urban driving conditions.  \\ \hline
Waymo Open Dataset \cite{2020_Sun_Kretzschmar_Dotiwalla} & 2020 & The Waymo Open Dataset features millions of 3D bounding boxes for vehicles, pedestrians, cyclists, and other objects encountered in urban and suburban environments. & 2,030 scenes for sensor data, and 103,354 scenes for motion data  \\ \hline
Lyft Level 5 Dataset \cite{2023_Li_Jiao} & 2021 & This dataset captured with a fleet of autonomous vehicles navigating through urban and suburban areas over a four-month period, enhanced with high-definition maps and traffic light statuses. & This dataset, consisting of over 170,000 scenes each lasting about 25 seconds, offering detailed perceptions and motion data.  \\ \hline
OpenACC \cite{2021_openACC} & 2021 & OpenACC dataset is collected for understanding of commercial ACC control logic and is designed for extensive research into vehicle performance and safety. & The dataset is a collection from 27 vehicle models across 17 brands, featuring a mix of engine types and captured using advanced GPS systems under various driving conditions.
\\ \hline
\end{tabularx}
\end{table*}




\section{CONCLUSION AND OUTLOOK}
This review aims to enhance understanding of the underlying design principles in different categories of CF Models, enabling researchers to formulate problems with a suitable framework. As the path to achieve level-5 autonomy continues, incorporating general traffic flow knowledge presents both challenges and opportunities. The future direction of driver behavior models, particularly in the context of integrating human-like reasoning capabilities and learning efficiencies into car following models, presents a fascinating and complex challenge. Bridging the gap between conventional fixed and myopic driver behavior models and rapidly advancing AI models requires a multifaceted approach. 

\textbf{Hybrid-learning Systems}: The potential of hybrid-learning systems lies in their ability to produce models that are not only highly predictive but also deeply understandable and controllable. Among the leading methodologies in this frontier is Physics-informed Machine Learning (PIML), which stands out for its ability to harmonize traditional kinematic models with the implicit understanding gleaned from vast datasets of real-world driving behaviors. PIML represents a significant step forward by embedding physical principles directly into machine learning frameworks, thereby ensuring that predictions not only fit observed data but also adhere to fundamental laws of motion. This alignment guarantees that the models remain valid under a wide range of conditions.

Another development in hybrid learning systems is Neural Symbolic AI, which seeks to bridge the gap between the intuitive pattern recognition capabilities of deep learning and the logical rigor of symbolic reasoning. By combining these two approaches, Neural Symbolic AI facilitates the construction of models that are both data-efficient and interpretable. These models excel in learning from available data while simultaneously encoding the structured knowledge that characterizes human reasoning processes. Such a dual approach enables the models to make sense of complex scenarios, predict outcomes, and make decisions that are both logically sound and grounded in real-world data. The inherent flexibility of Neural Symbolic AI allows it to adapt to new information and evolving scenarios, making it an ideal candidate for tackling the dynamic challenges of car-following behavior. As the field of autonomous driving continues to evolve, hybrid-learning systems stand poised to play a crucial role in developing models that seamlessly blend empirical data with theoretical insights, paving the way for the next generation of intelligent transportation systems that are both robust and adaptable.

\textbf{World Model Predictive Control}: The aspiration for driving agents to adapt swiftly to new environments and assimilate new tasks with near-human intelligence sets a high bar for the evolution of autonomous driving technologies. At the core of this evolution lies the concept of World Model Predictive Control, a paradigm shift that integrates common sense and a vast repository of background knowledge into the decision-making processes of next-generation intelligent driver models. This approach, inspired by advancements in Large Language Models (LLMs) like ChatGPT, represents a leap forward in enabling driving agents to comprehend and navigate complex driving scenarios with unprecedented depth and nuance.

The principle behind World Model Predictive Control is to endow driving agents with an internalized representation of the world. This internal model, enriched by extensive learning from diverse data sources, allows the agent to predict and plan several steps ahead, moving beyond reactive stimulus-response mechanisms to proactive situational management. Unlike traditional models that rely on immediate feedback or predefined control laws, this approach leverages predictive analytics to foresee and navigate future states, drawing on a comprehensive understanding of the driving environment's intricacies. The integration of such a world model into predictive control systems  harness the full power of artificial intelligence to ensure safety and efficiency on the roads.

\textbf{Hierarchical Planning and Control}: The ability to understand and apply abstract concepts is a defining feature of human intelligence. These frameworks structure decision-making across multiple layers, ranging from immediate sensory responses at the lowest level to strategic planning at the highest. Incorporating hierarchical learning into intelligent driving systems facilitates the encoding of diverse factors—such as vehicle characteristics, road conditions, and traffic dynamics—into high-level representations. This allows for the formulation of complex action plans that dynamically adapt to changing conditions, thereby enhancing operational safety and predictive accuracy. The intermediate levels of this hierarchy are particularly crucial for understanding and predicting the behaviors of nearby vehicles, improving the system's ability to anticipate and react to potential hazards before they arise. 

The integration of hierarchical planning and control into car-following models marks a significant step towards achieving true autonomy in driving technologies. By enabling vehicles to not only react to their immediate environment but also engage in anticipatory actions based on scenario analysis and predictive modeling, these systems enable agents to make decisions based on a blend of immediate data and abstract concepts. The development of such advanced models paves the way for a future where autonomous vehicles operate with a level of intuition and foresight akin to human drivers, contributing to safer, more efficient, and harmonious traffic ecosystems.

\textbf{Multimodal Signals}: The evolution of car-following models is increasingly pivoting towards the integration of multimodal signals to closely mimic the human driving experience, which inherently relies on a symphony of sensory inputs for decision-making. Current car-following models primarily utilize curated trajectory data, derived from post-processing, aggregation, and sensor fusion techniques. Human drivers base their decisions on a complex blend of visual cues, auditory signals, and even haptic feedback, something that the next generation of intelligent driving agents must emulate. By integrating information from a diverse array of sources—such as Cameras, LiDAR, Radar, Can-bus systems, Gyroscopes, and even Psychological Signals and Audio inputs—a richer, more precise understanding of the driving environment can be achieved. Approaches like multimodal fusion, which integrates data from these varied sources, cross-modality, which allows models to leverage information across different types of data, and shared representation learning, which finds commonalities across these data types, are crucial. This expanded use of multimodal signals promises to enhance the robustness, adaptability, and overall performance of car-following models, allowing them to better predict and react to the dynamic complexities of real-world driving.

\textbf{Self-Supervised Learning}: Traditional car-following models often rely heavily on specific objectives, predefined parameters, and extensively labeled data, which can limit their adaptability and applicability to new or unseen driving scenarios. Self-Supervised Learning (SSL) represents a paradigm shift, enabling models to learn from data that is not explicitly labeled for a specific prediction task. By exploiting the intrinsic structure of driving data, SSL techniques can uncover complex, latent relationships between past and future driving states without the need for direct supervision. This is achieved through mechanisms such as contrastive learning, which distinguishes between similar and dissimilar sequences of driving behavior, and regularized learning, which imposes additional constraints to guide the learning process. Moreover, SSL opens up new avenues for creating models that can adapt over time, learning from continuous streams of data generated by vehicles in operation. This ongoing learning process allows models to evolve in response to changing driving conditions and behaviors, significantly enhancing their long-term utility and effectiveness. The adoption of self-supervised learning in car-following models marks a critical step towards developing autonomous driving systems that can operate with an unprecedented level of sophistication and autonomy, mirroring the complex decision-making processes of human drivers.


\section{acknowledgement}
This paper is based upon work partially supported by National Science Foundation under Grant No. 1952096 and 2133516 and also by the U.S. Department of Homeland Security under Grant Award 22STESE00001-03-02 and 22STESE00001-01-01. The views and conclusions contained in this document are those of the authors and should not be interpreted as necessarily representing the official policies, either expressed or implied, of the U.S. Department of Homeland Security. 

\section{References}

\end{document}